\documentclass[superscriptaddress, twocolumn, amsmath, amssymb, aps, pra, letterdraft, notitlepage]{revtex4-1}
\usepackage[utf8]{inputenc}
\usepackage[english]{babel}
\usepackage[T1]{fontenc}
\usepackage{amsmath}
\usepackage{hyperref}
\usepackage{amsmath}
\usepackage{amssymb}
\usepackage{physics}
\usepackage{graphicx}
\usepackage{booktabs,multirow}
\hypersetup{
    colorlinks=true,
    linkcolor=blue,
    filecolor=blue,      
    urlcolor=blue,
    citecolor=blue,
}

\begin{document}
\title{Non-Markovianity Benefits Quantum Dynamics Simulation}

\author{Yu-Qin Chen}
\thanks{The two authors contributed equally to this work.} 
\email{yuqinchen@tencent.com}
\affiliation{Tencent Quantum Laboratory, Tencent, Shenzhen, Guangdong, 518057, China}

\author{Shi-Xin Zhang}
\thanks{The two authors contributed equally to this work.} 
\affiliation{Tencent Quantum Laboratory, Tencent, Shenzhen, Guangdong, 518057, China}

\author{Shengyu Zhang}
\email{shengyzhang@tencent.com}
\affiliation{Tencent Quantum Laboratory, Tencent, Shenzhen, Guangdong, 518057, China}

\begin{abstract}
Quantum dynamics simulation on analog quantum simulators and digital quantum computer platforms has emerged as a powerful and promising tool for understanding complex non-equilibrium physics. However, the impact of quantum noise on the dynamics simulation, particularly non-Markovian noise with memory effects, has remained elusive. In this Letter, we discover unexpected benefits of non-Markovianity of quantum noise in quantum dynamics simulation.
We demonstrate that non-Markovian noise with memory effects and temporal correlations can significantly improve the accuracy of quantum dynamics simulation compared to the Markovian noise of the same strength. 
Through analytical analysis and extensive numerical experiments, we showcase the positive effects of non-Markovian noise in various dynamics simulation scenarios, including decoherence dynamics of idle qubits, intriguing non-equilibrium dynamics observed in symmetry protected topological phases, and many-body localization phases. Our findings shed light on the importance of considering non-Markovianity in quantum dynamics simulation, and open up new avenues for investigating quantum phenomena and designing more efficient quantum technologies.
\end{abstract}

\maketitle
%Quantum systems always exist amidst environment noise. 
\textit{Introduction.---} Experiments conducted on artificially constructed quantum devices, designed for simulating quantum systems or performing quantum computations, are inevitably affected by environmental noise. Consequently, comprehending the nature of the noise environment and the intricate interplay between quantum systems and complex environments poses a fundamental scientific challenge. This field has attracted significant research interest from quantum chemistry, condensed matter physics~\cite{ashida2020non,bergholtz2021exceptional,heiss2012physics,lee2016anomalous,leykam2017edge,xu2017weyl,kunst2018biorthogonal,yao2018non,yao2018edge}, quantum information and computation~\cite{ingarden2013information,peres2004quantum,knill2005quantum,kitaev1997quantum}.

Most theoretical and experimental studies on quantum noise have been confined to the Markovian assumption, which posits that the dynamical evolution of a quantum system depends solely on its current state, disregarding its past history, and information continuously leaks into the environment over time. However, a more general scenario---the non-Markovian environment---is inevitable in most of the solid-state systems such as superconducting qubits, nitrogen-vacancy centers in diamond, and spin qubits in quantum dots~\cite{yoshihara2006decoherence,kakuyanagi2007dephasing,de2010universal,bar2013solid,kawakami2014electrical,watson2018programmable,pokharel2018demonstration,chen2020non}. For non-Markovian environments, the dynamical evolution of a quantum system depends on its current state as well as its past history, and the system information leaked into the environment can temporally flow back to the system, indicating a memory effect~\cite{breuer2016colloquium}. %Markovianity can be regarded as a simplified limit of non-Markovianity to zero-order memory effect. 
The general non-Markovian scenario is significantly more challenging for noise modeling and quantum dynamics analysis~\cite{breuer2002theory,rivas2010entanglement,breuer2009measure,cerrillo2014non,chen2020non,chen2022spectral} compared to the Markovian case, as the dynamical evolution is intimately affected by the entire past trajectory. 
Over the years, the influence of non-Markovianity on quantum information processing has been explored, including quantum Zeno effects~\cite{misra1977zeno},  continuous-variable quantum key distribution~\cite{vasile2011continuous},  quantum chaos~\cite{vznidarivc2011non}, quantum resource theory~\cite{wakakuwa2017operational}, and quantum metrology~\cite{matsuzaki2011magnetic,chin2012quantum}.

Despite the wealth of intriguing phenomena revealed under Markovian noise in physics systems and physics processes%including the interplay between non-Hermiticity and topology
~\cite{anderson1958random, Basko2006,bergholtz2021exceptional,heiss2012physics,lee2016anomalous,leykam2017edge,xu2017weyl,kunst2018biorthogonal,yao2018non,yao2018edge,zhang2021quench,lin2022experimental}, the interplay between non-Markovian noise and quantum dynamics simulation in physical systems remains elusive. Will the non-Markovianity lead to intriguing physics due to the environment's historical memory? How do memory effects impact the behavior of systems? What is the relationship between the impact on dynamical simulation under Markovian and non-Markovian baths? Addressing these questions is crucial for a comprehensive understanding of noise-induced effects in quantum systems, and potential utilization of non-Markovian noise for quantum information processing.

In this Letter, we bridge this gap and answer these important questions by investigating non-Markovian noise through the lens of quantum dynamics simulation in noisy environments. We discuss the general non-Markovian noise framework and characterize the dephasing noise with two key parameters identified and studied: noise strength and noise non-Markovianity. Our analysis encompasses various quantum dynamics simulation scenarios, including the decoherence dynamics of idle qubits, quench dynamics in symmetry-protected topological systems, and many-body localization systems. By systematically varying noise strength and non-Markovianity, we discovered that the dynamics simulation is more accurate and more closely resembles the clean dynamics with stronger non-Markovianity of the same noise strength under certain circumstances. In other words, non-Markovianity could \textit{mitigate} the detrimental effect of quantum noise, as opposed to the common conception that non-Markovianity always makes the quantum noise harder to cope with. This unexpected advantage offered by non-Markovianity in quantum noise holds great theoretical significance and experimental relevance for its potential to unlock new avenues for quantum engineering techniques and quantum error mitigation schemes. Our findings on the positive effects of non-Markovianity provide new insights for further exploration and utilization of non-Markovian noise in quantum systems.

\textit{Quantum dynamics in the non-Markovian environment.---} We briefly recapitulate the basic ingredients of non-Markovian noise and introduce two key parameters characterizing the pure dephasing noise investigated in this Letter with a simple idle qubit example.

We are concerned with a collection of qubits governed by the following Hamiltonian:
\begin{equation}
H(t)=H_s+\sum_i\textbf{B}_i(t)\cdot \boldsymbol{\sigma}_i,
\label{eq:totalHamiltonian}
\end{equation}
where $H_s$ is the time-independent system Hamiltonian.
For a quantum environment, $B_i^\alpha(t)=e^{-iH_bt}B_i^\alpha(0)e^{iH_bt}$ is a bath operator in the interaction picture with respect to the environmental Hamiltonian $H_b$. This bath operator can be regarded as a composition of real-valued stochastic processes when considering equivalent classical stochastic baths. Index $\alpha$  is one of the ${x,y,z}$ Cartesian components and index $i$ labels the qubits.
We consider the Gaussian noise, which is a reasonable assumption for most realistic scenes~\cite{erlingsson2002hyperfine,witzel2014converting,makri1999linear} and is fully characterized by two-time correlation function or the spectral density. By assuming that the initial system state $\rho(0)$ is independent from the environment, the time evolution for the system can be expressed in the following form:
\begin{equation}
\rho(t) = \langle\langle \exp_+(-i\int_0^t d\tau H(\tau)) \rho(0) \exp_{-}(i\int_0^t d\tau H(\tau)) \rangle\rangle_r,
\label{eq:evolution}
\end{equation}
where the subscripts $\pm$ on the exponential functions denote the (anti)chronological time ordering of the time-evolution operator, and $\langle \langle \cdots \rangle \rangle_{r}$ denotes an average over the environmental degrees of freedom. This dynamical process can be emulated by Monte Carlo trajectory averaged over stochastic process at equidistant time intervals, i.e., $t_n=n\delta t$, where the integer $n$ is chosen to be sufficiently large for convergence. %which ranges from zero to the total number of time steps. 
In this context, the corresponding unitary evolution operator from time $t_n$ to $t_{n+1}$ with one single noise configuration reads as $U(t_{n+1},t_n)=e^{-iH(t_n)\delta t}$. %The stochastic average of a physical operator $\mathbf{O}$ at time $t_n$ can be obtained by 
% \begin{equation}
% \langle \langle \langle \mathbf{O}\rangle\rangle\rangle_r=\Tr[\rho(t_n) \mathbf{O}].
% \end{equation}

It is a challenging task to figure out the general impact of stochastic baths on the dynamical process of arbitrary quantum systems. To begin with, we first examine the idle qubit case where analytical analysis is feasible. %For initially uncorrelated qubits and environments, 
The system state $\rho(0)$ is linearly mapped to the corresponding state $\rho(t)$ at time $t$ by a superoperator $\Lambda(t)$ called dynamical maps, i.e. $\rho(t)=\Lambda(t)\rho(0)$. For the Markovian environment, the dynamical map %preserves complete positive and trace-preserving (CPTP) and 
is divisible for all intermediate times $\tau$:
\begin{equation}
\Lambda_{t-0}=\Lambda_{t-\tau}\Lambda_{\tau-0}.
\label{eq:divisible}
\end{equation}
In the non-Markovian environment, however, Eq.~\eqref{eq:divisible} is violated in that the system at $t_n$ depends on the system's history through multiple transfer tensors with a memory 
effect~\cite{cerrillo2014non,chen2020non,chen2022spectral}:
\begin{equation}\label{eq:ttm}
\rho(t_n)=\sum_{m=1}^{K}T_m\rho(t_{n-m}),
\end{equation}
where $T_n \equiv \Lambda_n-\sum_m^{n-1}T_{n-m}\Lambda_m$, and $K=1$ corresponds to the Markovian limit. %The temporal correlation function of $\mathbf{B}_i(t)$ with finite correlation length can be recognized as the instigator of the non-Markovian memory effect. 
See the SM for the non-Markovian noise theory. 

In this Letter, we focus on pure dephasing noise. Specifically, we have $\langle \langle B_i^z(t) \rangle\rangle_{r}=0$ and the time correlation function is
\begin{eqnarray}\label{eq:auto1}
C(t-t')&\equiv& \langle\langle B_i^z(t)B_j^z(t')  \rangle\rangle_{r} \\ \nonumber
&=&\lambda e^{-b|t-t'|}\cos{[\omega_c(t-t')]}\delta_{i,j},
\end{eqnarray}
with $b^{-1}$ the time correlation length. Here $\delta_{i,j}$ indicates that the noises on different qubits are independent of each other.
Noise correlation as in Eq. \eqref{eq:auto1} corresponds to spectral density with Lorentzian profile via Fourier transformation~\cite{liu2011experimental,ma2014crossover},
\begin{equation}
S(\omega)=\frac{\lambda b}{b^2+(\omega-\omega_c)^2}.
\end{equation}
For the decoherence process of idle qubits, the dynamical map on each qubit can be analytically derived (see the SM for details),
\begin{equation} \label{eq:dymap}
\Lambda(t_n)=
\begin{bmatrix}
1 & 0 & 0 & 0 \\
0 & e^{-\Gamma(t_n)} & 0 & 0 \\
0 & 0 & e^{-\Gamma(t_n)} & 0 \\
0 & 0 & 0 & 1\\
\end{bmatrix},
\end{equation}
with $
\Gamma(t)=\frac{4}{\pi}\int_0^\infty d\omega \frac{S(\omega)}{\omega^2}[1-\cos{(\omega t)}]
$. It is straightforward to verify that the map in Eq.~\eqref{eq:dymap} is not divisible, i.e., $\Lambda_{n+m}\neq \Lambda_n \Lambda_m$, indicating non-Markovian behaviors. 

In order to quantitatively study the impact of non-Markovian effects while establishing a natural connection to results obtained under Markovian noise, we introduce two adjustable parameters: noise strength parameter and non-Markovianity parameter. Specifically, we denote $\lambda=C(0)$ in Eq.~\eqref{eq:auto1} as the noise strength parameter, consistent with the noise strength defined in widely studied Markovian Gaussian noise~\cite{zhang2021quench, lin2022experimental, Levi2016,wybo2020entanglement} sampled from $N(0,\sqrt{\lambda})$ with $\langle \langle B_i^\alpha(t) \rangle\rangle_{r}=0$ and $C(t-t')\equiv\langle \langle B_i^\alpha(t) B_j^\beta(t')\rangle\rangle_{r}=\lambda \delta(t-t')\delta_{\alpha,\beta}\delta_{i,j}$. Additionally, we denote $b$ in Eq.~\eqref{eq:auto1} as the non-Markovianity parameter, which governs the time correlation length and thus non-Markovian strength. When $b$ approaches infinity, the correlation function decays significantly faster than the characteristic time scale of the system's evolution and thus can be treated as a delta function leading to the Markovian limit.  Conversely, a smaller $b$ corresponds to a longer time correlation and stronger non-Markovianity. 
In the case of idle qubit dynamics as given in Eq. \eqref{eq:auto1}, a smaller $b$ results in a lower decoherence rate and thus dynamics with higher fidelity (see more details in the SM).

\begin{figure}[!htb]
\includegraphics[width=0.5\textwidth]{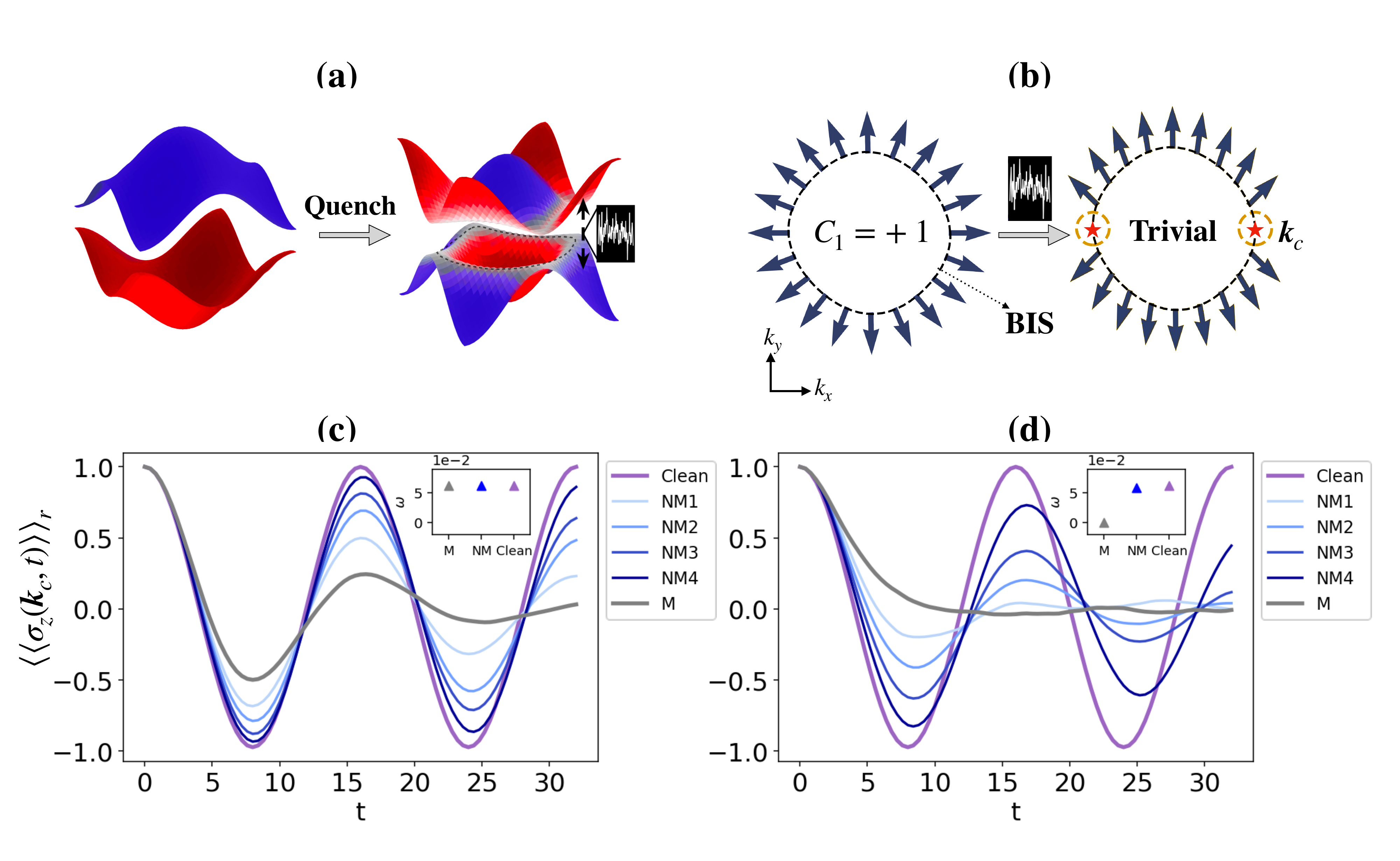}
\caption{(a) Topological quench dynamics influenced by quantum noise. (b) Quench-induced dynamical topology defined on BIS can be destroyed by sufficiently strong noise. $\boldsymbol{k}_c$ (red stars) denotes the singularity momentum on the dBIS, where the spin polarization dynamics lose oscillation behavior ($\overline{\langle\langle \boldsymbol{\sigma}(\mathbf{k})\rangle\rangle}_r \neq 0$), leading to ill-defined dBIS and the topological trivial phase. (c) The dynamical spin polarization of $\sigma_z$ on $\boldsymbol{k}_c$ under weak noise $\lambda=0.2 < \lambda_c$ ($\lambda_c\approx 0.39$ for studied model). (d) The dynamical spin polarization of $\sigma_z$ on $\boldsymbol{k}_c$ under strong noise $\lambda=0.8 > \lambda_c$.
In (c) and (d), the purple lines depict the results of the clean system, while the grey lines represent the results under Markovian noise. The blue lines labeled NM1, NM2, NM3, NM4 correspond to the results obtained with non-Markovian noise, with non-Markovianity parameter $b=2,1,0.6,0.2$, respectively. The insets in (c) and (d) display the oscillation frequency $\omega$ derived from the corresponding spin polarization dynamics.
} \label{fig:BIS2D}   % \label{}
\end{figure}

\textit{Quench-induced topological dynamics.---}Quantum quenches provide a nonequilibrium approach to investigating topological physics~\cite{vajna2015topological,caio2015quantum,budich2016dynamical,wilson2016remnant,flaschner2018observation,unal2020topological,hu2020topological}. This has spurred numerous experimental investigations across various quantum simulation platforms, including ultracold atoms \cite{sun2018uncover,yi2019observing}, nuclear magnetic resonance (NMR)~\cite{xin2020quantum}, nitrogen-vacancy defects in diamond \cite{ji2020quantum}, and superconducting circuits \cite{niu2021simulation}. Once referring to simulation experiments on quantum devices, environmental noise becomes an inevitable problem. Recently, there has been a surge of interest in exploring the interplay between topology physics and noisy environment, where diverse nonequilibrium topological phenomena emerged~\cite{heiss2012physics,lee2016anomalous,leykam2017edge,xu2017weyl,kunst2018biorthogonal,yao2018non,zhang2021quench,lin2022experimental}.

In this Letter, we explore the effect of non-Markovian noise in quench-induced dynamical topology simulation. We consider the 2D quantum anomalous Hall (QAH) model in momentum space,
\begin{equation}
H_{QAH}(\mathbf{k})=\mathbf{h}(\mathbf{k})\cdot \boldsymbol{\sigma},
\end{equation}
with Bloch vector $\mathbf{h}(\mathbf{k})=(t_{so}\sin{k_x},t_{so}\sin{k_y},m_z-t_0\cos{k_x}-t_0\cos{k_y})$. Here $t_0$ and $t_{so}$ are the spin-conserved and spin-flipped hopping coefficients, and $m_z$ is the magnetic field. We set the quenched Hamiltonian parameters as $m_z=1.2$, $t_{so}=0.2$, and $t_0=1$ throughout this work. For $0<|m_z|<2|t_0|$, $H_{QAH}$ exhibits nontrivial QAH behavior characterized by Chern number $C_1=-\text{sgn}\left(m_z\right)$. 

We focus on the quench dynamics  by varying $m_z$ suddenly from $m_z\gg t_0$ trivial phase ($\vert{\uparrow}\rangle$ for all $\mathbf{k}$) into $0<|m_z|<2|t_0|$ Chern insulating phase.
In this context, we define the dynamical spin polarization and the time-averaged spin polarization as
$\langle\langle \boldsymbol{\sigma}(\mathbf{k},t_n)\rangle\rangle_r=\text{Tr}[\rho(\mathbf{k},t_n)\boldsymbol{\sigma}]$
and  $\overline{\langle\langle \boldsymbol{\sigma}(\mathbf{k})\rangle\rangle}_r=\lim_{T\rightarrow{\infty}}\frac{1}{T}\int_0^T dt \ \langle\langle \boldsymbol{\sigma}(\mathbf{k},t)\rangle\rangle_r$, respectively. The nontrivial topology emerges on band inversion surface (BIS)~\cite{zhang2018dynamical,zhang2019dynamical,zhang2020unified,zhang2022universal} during the quench dynamics, where BIS is defined as a $\mathbf{k}$-subspace that supports spin-flip resonant oscillations, resulting in the vanishing time-averaged spin polarization $\overline{\langle\langle \boldsymbol{\sigma}(\mathbf{k})\rangle\rangle}_r=0$.%$\lim_{T\rightarrow{\infty}}\int_0^T dt \  \langle \psi(\mathbf{k},t) |\boldsymbol{\sigma}|\psi(\mathbf{k},t) \rangle|_{\mathbf{k}\in BIS}=0$. 

Once the QAH model is coupled to the stochastic environment, the Hamiltonian takes the form
\begin{equation}
H(\textbf{k},t)=H_{QAH}(\textbf{k})+\textbf{B}(\textbf{k},t)\cdot \boldsymbol{\sigma},
\label{eq:BIS}
\end{equation}
and the dynamics of the system are governed by the time evolution in Eq. \eqref{eq:evolution}.
Under Markovian noise~\cite{zhang2021quench,lin2022experimental}, the dynamical topology can be preserved on a noise-deformed band inversion surface (dBIS) if the noise strength is confined within a specific range, known as the sweet spot region. However, when the noise strength exceeds this sweet spot region, the bulk gap of the resulting dynamical phase closes. As a consequence, the dynamical oscillation of spin polarization at certain momentum vanishes, leading to an ill-defined dBIS and ultimately resulting in a topologically trivial phase.

We focus on pure dephasing bath defined in Eq. \eqref{eq:auto1}. %to study the noise effect on the quench-induced dynamical topology. 
In the Markovian limit, the sweet spot region that respects the dynamical topology is 
\begin{equation} \label{eq:region1}
\lambda \leq \lambda_c= \min_{\mathbf{k}\in \text{dBISs}} 2\|(h_x(\mathbf{k}),h_y(\mathbf{k}))\|,
\end{equation}
where $\lambda$ is the noise strength and $\lambda_c\approx 0.39$ in our setup. %and $h_x(\boldsymbol{k})=t_{so}\sin{k_x},h_y(\boldsymbol{k})=t_{so}\sin{k_y}$.
In terms of non-Markovian pure dephasing noise, the dynamics simulation results are presented in Fig.~\ref{fig:BIS2D}. When weak noise is within the sweet spot region (Fig.~\ref{fig:BIS2D}(c)), the dynamical spin polarization at $\boldsymbol{k}_c$ exhibits decay companing with oscillation behavior in the Markovian limit. After rescaling the dynamical process by neglecting the amplitude decay while keeping the oscillation, dBIS can be well defined ($\overline{\langle\langle \boldsymbol{\sigma}(\mathbf{k})\rangle\rangle}_r=0$) thereby preserving the dynamical topology. As non-Markovian effects are gradually introduced, the dynamical spin polarization dynamics approaches cleaner cases with weaker amplitude decay, compared to the original Markovian scenario.

For strong noise strength outside the sweet spot region in the Markovian limit (Fig.~\ref{fig:BIS2D}(d)), the dynamical spin polarization at $\boldsymbol{k}_c$ exhibits serious decay without oscillation behavior (oscillation frequency is $\omega=0$ given by Fourier transformation). Consequently, the dBIS becomes ill-defined, resulting in a noise induced topologically trivial phase. Once the non-Markovanity is turned on gradually, the dynamical spin polarization displays behavior similar to the weak noise case, with a finite oscillation frequency similar to the noiseless value. This leads to the successful recovery of a well-defined dBIS and the restoration of the topology from the detrimental effects of strong noise, i.e. non-Markovianity induced topological phase recovery. (See more numerical results in the SM).
%See more simulation results including dynamic spin polarization and time-average spin polarization in all spin directions, and entire Brillouin zones in SM). 
These results collectively demonstrate that the non-Markovianity presented in quantum noise benefits the dynamics simulation process and drives a topological phase recovering from a trivial one under strong noise.

\textit{Many-body localization dynamics.---}Many-body localization (MBL) is one of the cornerstones of non-equilibrium physics which brings a robust exception of eigenstate thermalization hypothesis \cite{Srednicki1994}. Starting from Anderson localization \cite{Anderson1958}, the system can avoid thermalization with many-body interactions turned on  \cite{Basko2006} due to the emergence of local integrals of motion \cite{Serbyn2013a}. Strong random disorder \cite{Oganesyan2007, Pal2010a, Nandkishore2015, Altman2014, Abanin2018}, quasiperiodic potential \cite{Iyer2013, Khemani2017, Zhang2018, Zhang2019a, Kohlert} and linear potential \cite{Schulz2019a, VanNieuwenburg2019, Khemani2020, Doggen2021a, Liu2022} can all lead to many-body localization in one dimension. As one of the most important non-equilibrium quantum phases as well as the foundation for other novel phases such as discrete time crystal \cite{Else2019, Zaletel2023} and Hilbert space fragmentation \cite{Khemani2020, Doggen2021a, Yang2020b}, MBL system has been extensively explored on different experimental platforms of programmable quantum simulators and quantum computers \cite{Schreiber2015a, Smith2016, Liu2021c, Gong2021}. Therefore, a comprehensive investigation of the quantum noise effect for MBL simulation is highly desired. The decoherence effect on the MBL system has been investigated in Ref. \cite{Levi2016}, where only Markovian dephasing noise is considered.% while the effect of non-Markovian noise on MBL phase remains elusive.

In this Letter, we investigate the interplay between quantum decoherence with and without non-Markovianity and the one-dimensional spin model hosting MBL with on-site quenched disorder. The MBL Hamiltonian with open boundary conditions reads: 
\begin{equation}
H_{\text{MBL}} = \sum_{i}\left ( S^x_{i}S^x_{i+1} + S^y_{i}S^y_{i+1} + J_z S^z_{i}S^z_{i+1} \right )+ \sum_i W_i S^z_i.
\label{eq:MBL}
\end{equation}
The interaction term $J_z=0.2$, and $W_i$ is the static-random fields sampled from a Gaussian distribution $N(0,W/\sqrt{3})$, where we use $W=10$ deeply in the MBL phases. Note that the random variable is so chosen that it can be directly comparable with the form of dephasing noise (see the SM for details). The phase diagram for this clean model can also be found in the SM.

We employ two representative observables: charge imbalance and entanglement entropy to probe the characteristics of MBL dynamics. Charge imbalance $\mathcal{I}$ is defined as the difference in occupation numbers between odd and even sites $\mathcal{I} = \sum_{i}(-)^i\langle S^z_i\rangle$, assuming that the initial state is an antiferromagnetic product state. %This quantity is straightforward and commonly used to observe MBL in experiments \cite{Schreiber2015a}. 
It is expected to be (non-)zero for thermal (MBL) phases at infinite time limit with exponential decay at later times in thermal phases. The half-chain entanglement entropy $S$, defined as the von Neumann entropy of the reduced density matrix on half of the system, provides a more intrinsic information perspective of MBL phases, exhibiting a logarithmic increase before saturating to a non-thermal value \cite{Bardarson2012, Abanin2013}. In systems of very weak thermalization, e.g. MBL systems coupled to weak Markovian dephasing \cite{Levi2016,wybo2020entanglement}, entanglement can still follow the logarithmic scaling but with saturating value $S(t\rightarrow \infty)$ the same as the typical thermal value. 

\begin{figure}[!htb]
\includegraphics[width=0.5\textwidth]{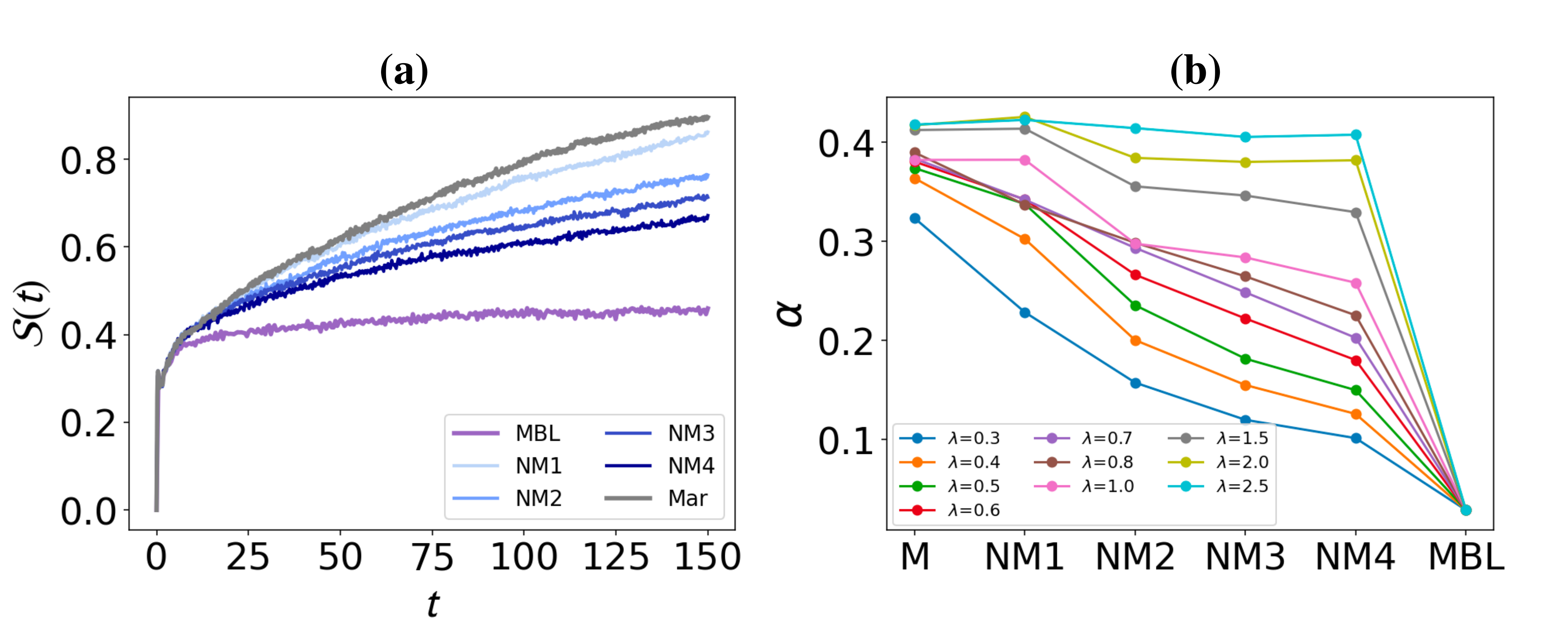}
\caption{(a) Entanglement entropy of many-body localization dynamics with pure dephasing noise of strength $\lambda=0.4$. The purple line shows the results of
the clean MBL system. The grey line shows the results under Markovian noise. Blue lines labeled from NM1 to NM4 show the results under non-Markovian noise with non-Markovianity parameter $b =  4, 1, 0.5, 0.2$. (b) Parameter $\alpha$ that fits the late-time entanglement entropy dynamics in scaling relation $S(t)\sim \alpha \log(t+\gamma)+\beta$ under Markovian bath (M), non-Markovian bath (NM1-NM4) and original MBL with varying noise strength $\lambda$.
} \label{fig:ee}  
\end{figure}

\begin{figure}[!htb]
\includegraphics[width=0.5\textwidth]{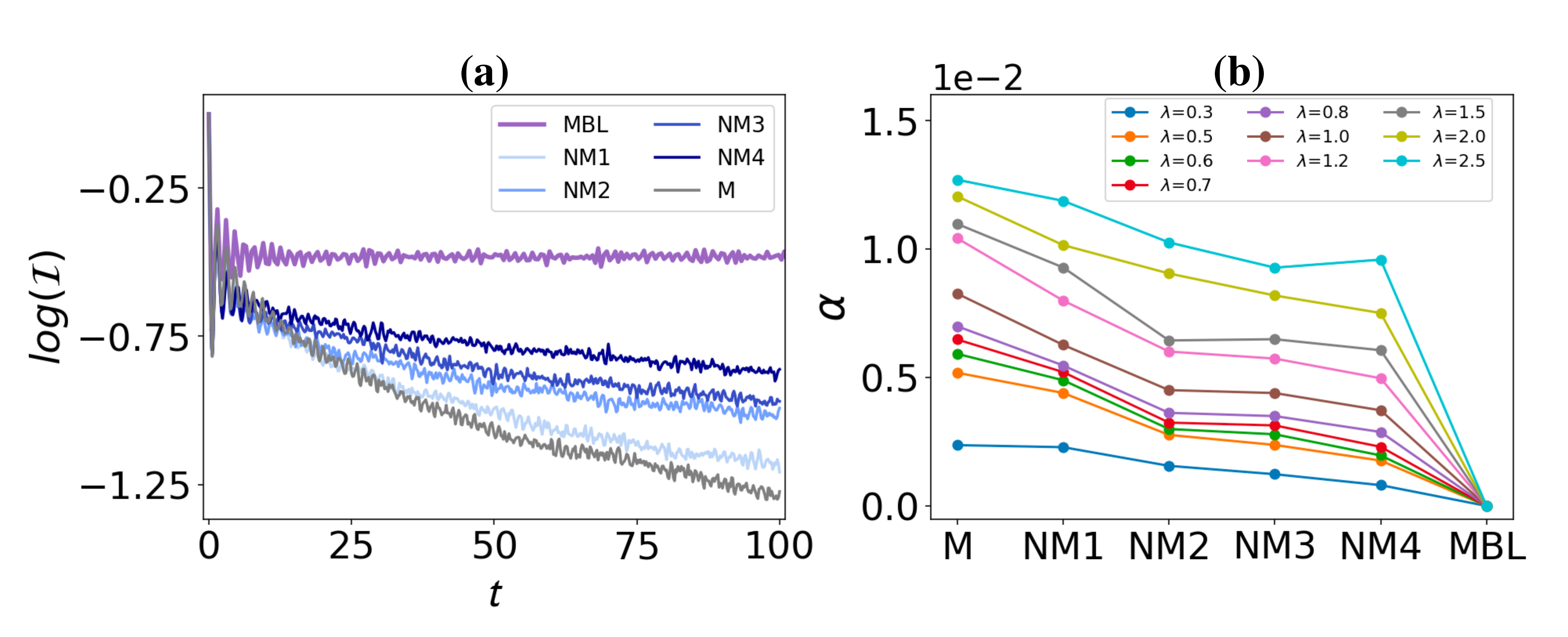}
\caption{(a) Charge imbalance of many-body localization dynamics with pure dephasing noise of strength $\lambda=0.7$. The purple line shows the results of
the clean MBL system. The grey line shows the results under Markovian noise. Blue lines labeled from NM1 to NM4 show the results under non-Markovian noise with non-Markovianity parameter $b =  4, 1, 0.5, 0.2$. (b) Parameter $\alpha$ that fits the late-time charge imbalance dynamics in scaling relation $\mathcal{I}(t)\sim e^{-\alpha t+\beta}$ under Markovian noise (M), non-Markovian noise (NM1-NM4) and clean MBL with varying noise strength $\lambda$.
} \label{fig:ci}   
\end{figure}

To compute the entanglement entropy in the context of Monte Carlo trajectory simulation for the quantum noise environment, we compute the trajectory averaged entanglement entropy but not the reverse (entropy on trajectory averaged density matrix). The conceptual subtlety here is from the non-linearity of the definition of entropy \cite{Skinner2019a, Chan2019, Li2018b}. In view of trajectory level dynamics, MBL dynamics with single noise realization introduce temporal randomness apart from the spatial randomness presented in pure MBL system. The difference is similar to the Floquet circuit/random circuit or static Hamiltonian/Brownian Hamiltonian \cite{Lashkari2013, Sahu2023} scenarios. In our setup, dephasing noises with and without non-Markovianity enter the external field $W_i$ term in the Hamiltonian Eq. \eqref{eq:MBL}. The Markovian quantum noise is independent at each time slice and thus is of Brownian type. On the contrary, non-Markovian noise has a temporal correlation between different times and goes beyond the Brownian paradigm.

With the introduction of quantum noise in the time evolution, i.e. with the coupling to the environment, the system under investigation will eventually be thermal. One interesting question that arises is whether the non-Markovianity in quantum noise can lead to slower thermalization and preserve the many-body localization characteristics to some extent in the simulation. We numerically study the MBL dynamics simulation of $N=8$ qubit system with (non-)Markovian quantum noise of different strengths and different non-Markovianity degrees. The results for entanglement entropy and charge imbalance are illustrated in Fig.~\ref{fig:ee} and Fig.~\ref{fig:ci}, respectively. The left panel of both figures shows the dynamics of a given noise strength $\lambda$ with varying non-Markovianity. In the right panel, we fit the later time dynamics to extract the relevant scaling quantities reflecting the speed of thermalization. For late-time dynamics, we use the scaling relation $\alpha \log (t+\gamma)+\beta$ and $e^{-\alpha t+\beta}$ to fit these two quantities. In both cases, a larger positive $\alpha$ reflects a quicker thermalization. The results in panel (b) of both figures demonstrate that larger noise strength $\lambda$ can lead to stronger thermalization. More interestingly, given the same strength of the quantum noise, the stronger non-Markovianity can lead to a weaker thermalization, indicating a better reflection of the clean MBL dynamics. In other words, the memory effect rooted in non-Markovianity makes it harder to erase the initial state information, which is a signal against thermalization.

\textit{Conclusion.---} In this Letter, we have investigated the impact of non-Markovian quantum noise on quantum dynamics simulations using both analytical and numerical approaches \cite{Zhang2022}. Remarkably, our findings indicate that non-Markovianity can have a beneficial effect by offering enhanced protection for clean dynamics compared to noise of equivalent strength in the Markovian limit.
This unexpected conclusion highlights the potential advantages of incorporating non-Markovian noise in quantum simulations. By better preserving the integrity of the system's dynamics, non-Markovianity offers a promising avenue for improving the accuracy and reliability of quantum simulations across diverse quantum platforms.

\bibliographystyle{apsreve}
\bibliography{manuscript}
\clearpage
\newpage

% \appendix
\onecolumngrid
\renewcommand{\thesection}{S\arabic{section}}
\setcounter{section}{0}
\renewcommand{\theequation}{S\arabic{equation}}
\setcounter{equation}{0}
\renewcommand{\thefigure}{S\arabic{figure}}
\setcounter{figure}{0}
\renewcommand{\thetable}{S\arabic{table}}
\setcounter{table}{0}

\section*{Supplemental Materials \\Non-Markovianity Benefits Quantum Dynamics Simulation}

\section{Time-nonlocal Quantum Master Equation} \label{sec:time-nonlocal}
In this section, we briefly introduce the time-nonlocal master equation that describes general open quantum systems in terms of both quantum environment and classical stochastic noise.

Once a time-independent system $H_s$ couples to stochastic baths, the Hamiltonian can be described as
\begin{equation}
H(t)=H_s+H_{sb}=H_s+\sum_i\textbf{B}_i(t)\cdot \boldsymbol{\sigma}_i,
\end{equation}
where $H_s$ is the time-independent system Hamiltonian.
For quantum noise, $B_i^\alpha(t)=e^{-iH_bt}B_i^\alpha(0)e^{iH_bt}$ is a bath operator in the interaction picture with respect to $H_b$ -- the environment Hamiltonian. Index $\alpha$  is one of the ${x,y,z}$ Cartesian components and index $i$ labels the freedoms of qubits. For classical stochastic baths, the system is not explicitly coupled to another quantum system but is subjected to classical noise sources, i.e., $B_i^\alpha(t)$ is made up of real-valued stochastic processes. 

The dynamical process of general open quantum systems is described by the time-nonlocal quantum master equation (TNQME)~\cite{breuer2002theory}:
\begin{equation}\label{eq:TNQME}
\frac{d\rho(t)}{dt}=\mathcal{L}_s\rho(t)+\int_0^tdt'\mathcal{K}(t-t')\rho(t'),
\end{equation}
where $\rho(t)$ denotes the system's reduced density matrix, $\mathcal{L}_s$ is the Liouville superoperator corresponding to the system Hamiltonian, and $\mathcal{K}(t)$ is a memory kernel that fully encodes environment-induced decoherence effects and the memory effect of the evolution
history from non-Markovian effect.  In terms of the Markovian approximation, the memory kernel is local in time $K(t-t')\propto \delta(t-t')$ and the master equation is reduced into the Lindblad form. It is also reported that~\cite{chen2020non}, under the proper definition, dynamics induced by classical stochastic  noise can be described by a formally identical time-nonlocal master equation as the quantum one in Eq.~\eqref{eq:TNQME}. 

We consider the Gaussian noise, which can be fully characterized by the first two statistical moments $\langle B_i^\alpha(t)\rangle$ and a two-time correlation function:
\begin{eqnarray} \label{eq:cc}
C_{\alpha\alpha'}(t) = \langle B^{\alpha}(t)B^{\alpha'}(0) \rangle.
\end{eqnarray}
For Gaussian noise, the memory kernel $\mathcal{K}(t)=\sum_{n=1}^\infty \mathcal{K}_{2n}(t)$, where $K_{2n}(t)$ denotes order-$2n$ expansion of the Hamiltonian with respect to $H_{sb}$. The memory kernel is directly related to time correlation function of the noise profile. For example, the leading order of memory kernel in a weak noise regime gives
\begin{eqnarray} \label{eq:ke}
\mathcal{K}(t)(\cdot) \approx \sum_{\alpha,\alpha'}[\sigma^\alpha,C_{\alpha \alpha'}(t)\sigma^{\alpha'}(t)(\cdot)-C_{\alpha\alpha'}^*(t)(\cdot)\sigma^{\alpha'}(t)],
\end{eqnarray}
where $\sigma^\alpha(t)=e^{-iH_st}\sigma^\alpha e^{iH_st}$ and $C_{\alpha\alpha'}(t)$ is the time correlation function defined in Eq.~\eqref{eq:cc}. Namely, we can replace the quantum bath with classical stochastic noise with the identical correlation matrix and obtain the same noisy dynamics. This equivalence underlines the Monte Carlo trajectory noisy dynamics numerical simulation of this work. %For classical stochastic noise, the memory kernel retains the functional form  in Eq.~\eqref{eq:ke} with real-valued $C_{\alpha\alpha'}(t)&=& \langle\langle B^\alpha(t)B^{\alpha'}(0) \rangle\rangle_{r}$, and $\langle\langle \cdots\rangle\rangle_r$ denotes the average of different noise configurations. 
The noise spectral density is defined for real valued correlation matrix via Fourier transformation as
\begin{equation}
S_{\alpha\alpha'}(\omega)=\int_{-\infty}^{\infty}dt e^{i\omega t}C_{\alpha\alpha'}(t).
\end{equation}

If the correlation function for the environment noise decays significantly faster than the characteristic time scale of the system's evolution, the temporal correlation function can be treated as a delta function with a uniform spectral density. The correlation time and the memory effect in this Markovian limit are regarded to be zero. Therefore, Markovian noise is only a special limit of the general non-Markovian noise.

\section{Non-Markovian Pure-dephasing Noise} \label{sec:appnmpure}
In this section, we present the theory framework of non-Markovian pure-dephasing noise and analytically analyze corresponding dynamical behavior induced by non-Markovianity for idle qubits.

In general, decoherence coming from the environment is a complicated process involving both population relaxation and dephasing. However, when we enhance the intensity of an external polarization field to a point where the energy level separations are significantly large, the relaxation process can be rendered insignificant within the relevant timeframe, leaving dephasing as the sole main cause of decoherence. In the context of this pure dephasing assumption, the environmental noise can be represented as random fields that induce random phase shifts. This approach is applicable in various dephasing environments, such as spin baths in solid-state systems~\cite{biercuk2009optimized,de2010universal,ajoy2011optimal} and background noise in superconducting qubits~\cite{cywinski2008enhance}. The pure dephasing Hamiltonian of $N$-qubit open systems reads 
\begin{equation}
H_{sb}=\sum_{i=1}^N B_i^z(t)\sigma_{z_i},
\label{eq:pure}
\end{equation}
where we assume the Gaussian noise with  $\langle\langle B_i^z(t) \rangle\rangle_r =0$ and time correlation function $C(t-t')=\langle\langle B_i^z(t) B_j^z(t')\rangle\rangle_r\delta_{i,j}$. $\delta_{i,j}$ denotes that the noise experienced by each qubit is independent of each other. The temporal correlation function of
$B_i(t)$ with finite correlation length can be recognized as the instigator of the non-Markovian memory effect, see Sec.~\ref{sec:time-nonlocal}.

Considering a general initial pure state $|\psi(0)\rangle= \sum_{\vec{s}=00\cdots 0}^{11\cdots 1}\alpha_{\vec{s}}|\vec{s}\rangle$, $\vec{s}=s_1s_2\cdots s_N$, $s_i=0$ or $1$, the quantum state at time $t$ under pure dephasing bath becomes
\begin{eqnarray}
\bar{\rho}_{\vec{s},\vec{s}'}(t)=\alpha_{\vec{s}}\alpha_{\vec{s}'}^*\langle\langle e^{-2i\sum_{l=1}^N (s'_l-s_l)F_l(t)} \rangle\rangle_r
\end{eqnarray}
where $F_l(t)\equiv \int_0^t B_l(\tau)d\tau$ and $\langle\langle \cdots \rangle\rangle_r$ denotes average over noise configurations.
For initially uncorrelated qubits and environments, the open system state $\rho(0)$ is linearly mapped to the corresponding state at time $t$ by dynamical maps $\Lambda(t)$
\begin{equation}
\rho(t)=\Lambda(t)\rho(0)
\end{equation}
Dynamical maps $\Lambda$ are superoperators (of dimension $d^2\times d^2$, $d=2^N$) and the map is carried out as a matrix-vector product when the density matrix $\rho$ is vectorized as length $d^2$. The structure form of $\Lambda$ brings many operational advantages. For instance, a succession of transitions can be performed with simple matrix multiplications, eg., $\Lambda_1\Lambda_2\rho$.
For stochastic noise that is independent between qubits, we can focus on single-qubit dynamical maps, and the multiple-qubit dynamical maps can be regarded as tensor products of single-qubit ones. More complicated decoherence coming from correlated noise, i.e., $\langle\langle B^z_i(t)B^z_j(t')\rangle\rangle_r\neq 0$ for $i\neq j$ was presented in Ref.~\cite{chen2020non}. The single-qubit dynamics in terms of density matrix elements reads
\begin{eqnarray}\label{eq:rho} 
\bar{\rho}_{0,1}(t)&=&\alpha_0\alpha_1^*\langle\langle e^{-2iF(t)}\rangle\rangle_r =\bar{\rho}_{0,1}(0) e^{-2\langle\langle F(t)^2\rangle\rangle_r}=\bar{\rho}_{0,1}(0)e^{-\Gamma(t)}, \\  \nonumber
\bar{\rho}_{1,0}(t)&=&\alpha_0^*\alpha_1\langle\langle e^{-2iF(t)}\rangle\rangle_r =\bar{\rho}_{1,0}(0) e^{-2\langle\langle F(t)^2\rangle\rangle_r}=\bar{\rho}_{1,0}(0)e^{-\Gamma(t)}, \\  \nonumber
\bar{\rho}_{0,0}(t)&=&\bar{\rho}_{0,0}(0), \  \bar{\rho}_{1,1}(t)=\bar{\rho}_{1,1}(0),
\end{eqnarray}
where the decoherence rate $\Gamma(t)$ can be obtained from time correlation function
\begin{eqnarray} \label{eq:gamma}
\Gamma(t)&=&2\langle\langle F(t)^2\rangle\rangle_n \\ \nonumber
&=&2\int_0^t\int_0^t \langle\langle B^z(t') B^z(t'')\rangle\rangle_n dt' dt'' \\ \nonumber
&=&2\int_0^t\int_0^t C(t'-t'') dt' dt''
\end{eqnarray} 
The noise spectral density $S(\omega)$ provides the energy distribution of the noise signal at different frequencies and often be used as the most important characteristic of noise sources. In this work, we assume the reservoir has a Lorentzian spectrum~\cite{imamog1994stochastic,park2000intraband},
\begin{equation}
S(\omega)=\frac{\lambda b}{b^2+(\omega-\omega_c)^2}.
\end{equation}
According to Sec.~\ref{sec:time-nonlocal}, the corresponding time correlation function becomes
\begin{equation}
C(t-t')=\lambda e^{-b|t-t'|} \cos{[\omega_c(t-t')]}.
\label{eq:correlation} 
\end{equation}
Based on Eq.~\ref{eq:rho} and Eq.~\ref{eq:gamma}, we can analytically obtain the pure-dephasing dynamical maps,
\begin{equation}
\Lambda(t)=
\begin{bmatrix}
1 & 0 & 0 & 0 \\
0 & e^{-\Gamma(t)} & 0 & 0 \\
0 & 0 & e^{-\Gamma(t)} & 0 \\
0 & 0 & 0 & 1\\
\end{bmatrix},
\label{eq:map}
\end{equation}
with 
\begin{equation}
\Gamma(t)=\frac{4}{\pi}\int_0^\infty d\omega \frac{S(\omega)}{\omega^2}[1-\cos{(\omega t)}]. 
\end{equation}
It is straightforward to verify that these maps are in general not divisible, i.e.,$\Lambda_{n+m}\neq \Lambda_n \Lambda_m$. We note that those analytical results benefiting our understanding are owing to the pure dephasing case in which all terms in $H_{sb}$ commute with each other. Otherwise, the dynamical maps and non-Markovianity have to be obtained by various quantum tomography methods~\cite{chen2020non,chen2022spectral}.  

We denote $C(0)=\lambda$ in Eq.~\eqref{eq:correlation} as the noise strength parameter, which can be directly interpolated to the noise strength defined in the usual Markovian bath. For instance, for white noise sampled from Gaussian distribution $N(0,\sqrt{\lambda})$, the correlation function exhibits $C(t)=\lambda\delta(t)$ with noise strength $\lambda$. Additionally, $b$ in Eq. \eqref{eq:correlation} characterizes the non-Markovianity, which governs the correlation time scale for the noise profile. With smaller $b$, the correlation length is larger, indicating stronger non-Markovianity. We discuss cases with different values of $b$:

1) Markovian limit $b\rightarrow \infty$,
\begin{eqnarray}
|C(t-t')|=\lambda \cdot \delta(t-t').
\end{eqnarray}
For pure dephasing in Eq.~\eqref{eq:map}, $\Gamma(t)=2\lambda t$, the dynamical map satisfies $\Lambda_{n+m}= \Lambda_n \Lambda_m$, and is thus Markovian.

2)Dynamical noise processes with finite $b$ lead to moderate non-Markovianity.

\begin{figure}[!htb]
\includegraphics[width=0.95\textwidth]{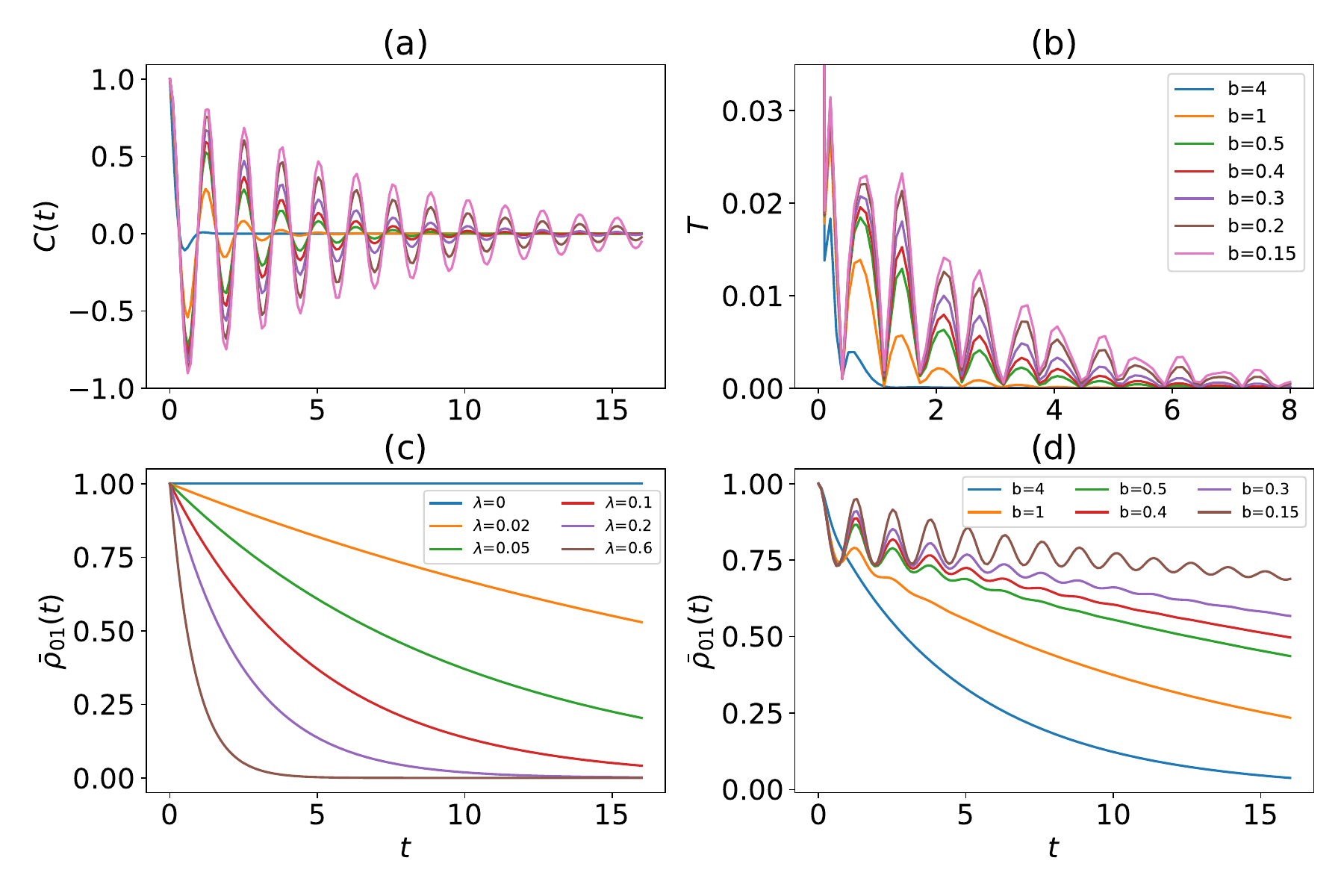}
\caption{Non-Markovian pure dephasing noise in Lorentzian form. (a) The real part of correlation functions changes along with time with $\lambda=1$. (b) The transfer tensor maps in 2-norm change along with time distance $t=\delta t \,n$ ($\delta t=0.2$). (c) The relaxation dynamics of the idle qubit under varying noise strength $\lambda$ with $b=1$. (d) The relaxation dynamics of the idle qubit under varying non-Markovianity $b$ with $\lambda=1$. We set $\omega_c=5$. 
} \label{fig:nonmar}   % \label{}
\end{figure}

According to Sec.~\ref{sec:time-nonlocal}, we qualitatively realize that the temporal correlation function of $B_i(t)$ with finite correlation time can be recognized as the instigator of the non-Markovian memory effect. We further involve transfer tensor to quantitatively represent the relationship between correlation function and non-Markovianity. Dynamical evolution up to time $t_n$ depends on the system’s history through multiple transfer tensors indicating a memory effect
\begin{equation}
\rho(t_n)=\sum_{m=1}^{K}T_m\rho(t_{n-m}), \ \  
\end{equation}
which offers a simple discretized way to characterize non-Markovian evolution since $T_n\approx \mathcal{K}(t_n)\delta t^2$. 
More effective transfer tensor maps (larger $K$, $\| T_i \|\geq\epsilon, i \in [1,K]$) denote stronger non-Markovianity and longer history memory.
We show it explicitly with transfer tensor maps stated in the main text in Fig.~\ref{fig:nonmar}. As we can see, smaller $b$ corresponds to a longer correlation length in Fig.~\ref{fig:nonmar}(a) and stronger non-Markovianity (longer memory on the history dynamics) in Fig.~\ref{fig:nonmar}(b).

Starting from an initial state $|\psi(0)\rangle=|0\rangle+|1\rangle$, according to Eq.~\eqref{eq:map}, the instantaneous state can be derived, see Fig.~\ref{fig:nonmar}(c)(d). It is obvious that the decoherence of the initial state is weaker under smaller noise strength $\lambda$ with the same $b$ and under stronger non-Markovianity (smaller $b$) with the same $\lambda$.

\section{Stochastic Simulation of non-Markovian Processes} \label{sec:appnmsim}
In the former section, we analytically analyze the dynamical maps and non-Markovian behavior focusing on the pure dephasing noise bath alone ($H_s=0$).
Once we want to simulate interesting quantum systems $H_s$ in a general non-Markovian reservoir, i.e.,$H=H_s+H_{sb}$, the quantitative consequence is hard to track analytically even in the non-Markovian pure dephasing noise case. Instead, we simulate the dynamics by stochastic processes in discretized time steps, in which the stochastic noise is generated using the Fast Fourier Transform (FFT) method~\cite{heideman1985gauss}. The average results over different stochastic processes are equivalent to the dynamics of the given system with non-Markovian noise.

This numerical method uses the relation of the correlation function $C(t)$ to the non-negative real-valued spectral density  $S(\omega)$.
The integral can be approximated by a discrete integration scheme
\begin{equation}
C(t) = \int_{-\infty}^{\infty} \mathrm{d}\omega \, \frac{S(\omega)}{\pi} e^{-\mathrm{i}\omega t}
\approx \sum_{k=0}^{N-1} \Delta\omega \frac{S(\omega_k)}{\pi} e^{-\mathrm{i} \omega_k t}.
\end{equation}
where  $\omega_k=\omega_{min}+k\Delta \omega$, $\Delta \omega=(\omega_{max}-\omega_{min})/(N-1)$, $t_n=n\delta t$ and $N\Delta \omega \delta t=2\pi$. $N,\omega_{min},\omega_{max}$ are hyperparameters set to satisfy the required tolerance from the following approximations.   

For a stochastic process defined as 
\begin{equation}
f(t) = \sum_{k=0}^{N-1} \sqrt{\frac{\Delta\omega S(\omega_k)}{\pi}} Y_k e^{-\mathrm{i}\omega_k t},
\end{equation}
with independent complex random variables $Y_k$ such that $\langle Y_k \rangle = 0$ and $\langle Y_k Y^\ast_{k'}\rangle = \delta_{k,k'}$,
it is easy to identify that its correlation function will be the same as the expected correlation function from non-Markovian dephasing noise,
\begin{eqnarray}
\langle f(t) f^\ast(t') \rangle&=& \sum_{k,k'} \frac{1}{\pi} \sqrt{\Delta\omega^2 S(\omega_k)S(\omega_{k'})} \langle Y_k Y^\ast_{k'}\rangle \exp[-\mathrm{i}(\omega_k t - \omega_{k'} t')] \\ \nonumber
&=& \sum_{k}    \frac{\Delta\omega}{\pi} S(\omega_k) e^{-\mathrm{i}\omega_k (t-t')} \\ \nonumber
&\approx& C(t-t').
\end{eqnarray}
The discrete-time stochastic process of $f(t)$ in the time domain can be realized by
\begin{equation}
f(t_n) = e^{-\mathrm{i}\omega_\mathrm{min} t_n} \mathrm{FFT}\left( \sqrt{\frac{\Delta\omega S(\omega_k)}{\pi}} Y_k \right),
\end{equation}
where $Y_k$ are Gaussian distributed random variables.
It is easy to see that, the discrete-time stochastic process generated above can support noise in different random field directions and even noise correlation between qubits, i.e., $B^\alpha(t), \alpha=x,y,z$ or $\langle B_iB_j\rangle _{i\neq j}\neq 0$ beyond the qubit-independent pure dephasing model in Sec.~\ref{sec:appnmpure}.

\section{Non-Markovianity in quench-induced Dynamical Topology Simulation} \label{sec:appbis}
\begin{figure}[!htb]
\includegraphics[width=0.95\textwidth]{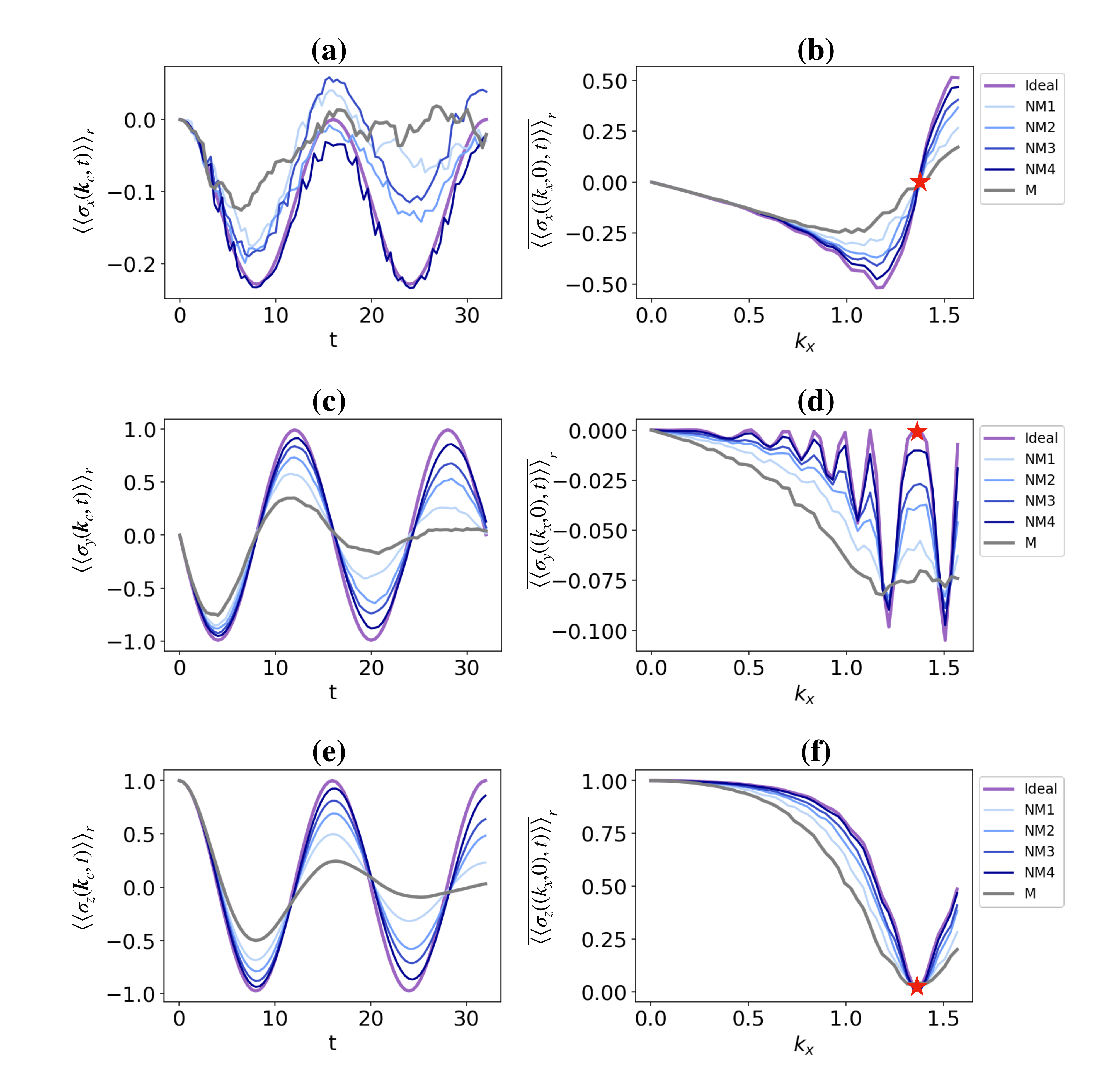}
\caption{Dynamical spin polarization and time-averaged spin polarization. Under weak noise strength $\lambda=0.2 < \lambda_c$ ( $\lambda_c\approx0.39$ for the studied model): The dynamical spin polarization (a)$\langle\langle \sigma_x \rangle\rangle_r$, (c) $\langle\langle \sigma_y \rangle\rangle_r$, (e) $\langle\langle \sigma_z \rangle\rangle_r$ on dBIS at $\boldsymbol{k}_c$. $\boldsymbol{k}_c$ (red stars) denotes the singularity momentum (defined in the main text or the right plane of this figure), the spin polarization dynamics first lose oscillation on the dBIS at $\boldsymbol{k}_c$ with the increase of noise leading to ill-defined dBIS and topological trivial phases. The time averaged spin polarization (b) $\overline{\langle\langle \sigma_x \rangle\rangle}_r$, (d) $\overline{\langle\langle \sigma_y \rangle\rangle_r}$, (f) $\overline{\langle\langle \sigma_z\rangle\rangle_r}$ at the momentum line $k_y=0$.  The purple lines (Ideal) show the results of a clean system. The grey lines (M) show the results under Markovian noise. The blue lines from lightest to heaviest (NM1,NM2,NM3,NM4) 
 show the results of non-Markovian noise with non-Markovianity parameterc $b = 2, 1, 0.6, 0.2$. The system parameters are chosen as $m_z=1.2,t_{so}=0.2,t_0=1$.
} \label{fig:2Dweak}   % \label{}
\end{figure}

\begin{figure}[!htb]
\includegraphics[width=0.95\textwidth]{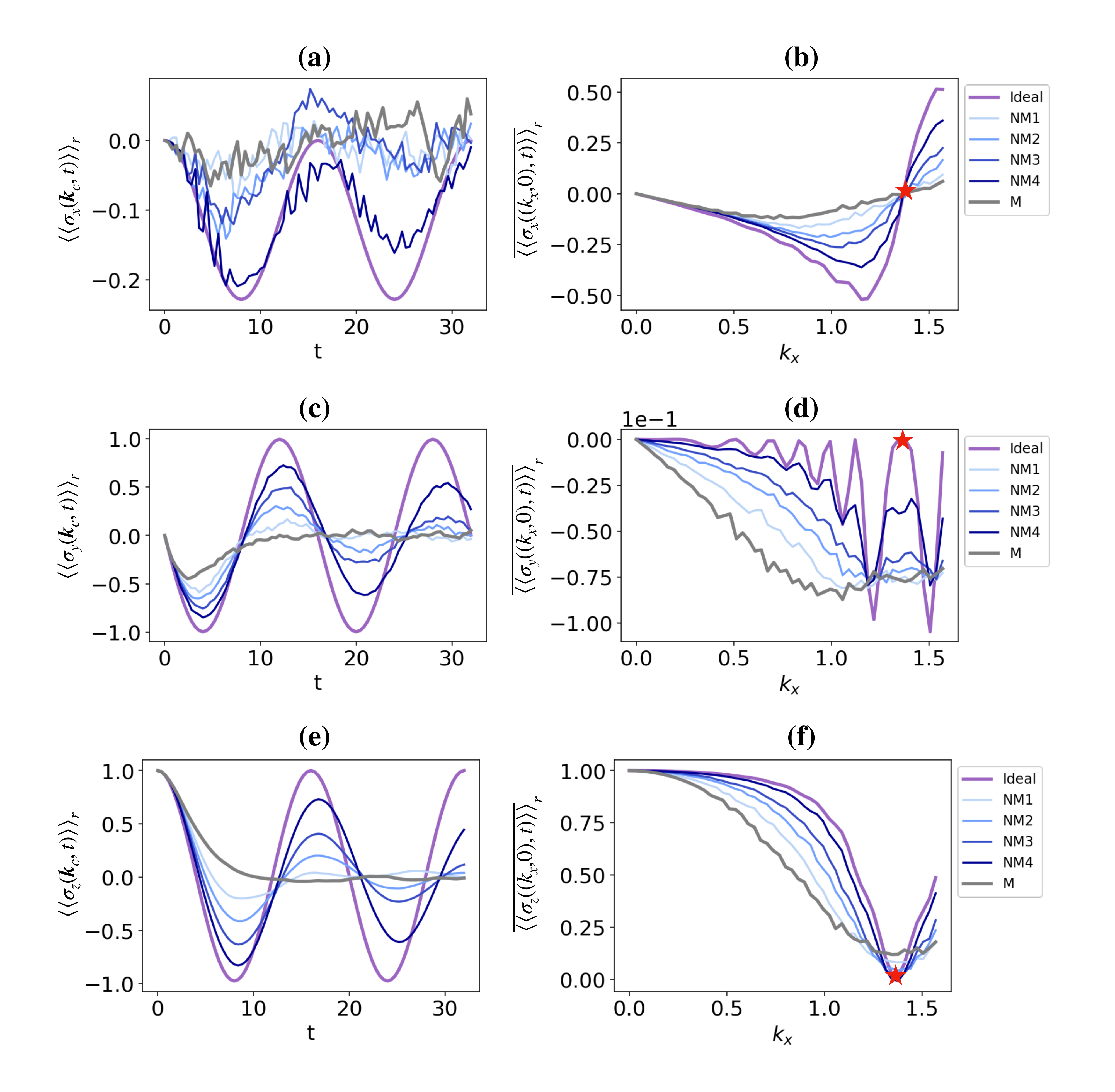}
\caption{Dynamical spin polarization and time-averaged spin polarization. Under strong noise strength $\lambda=0.8 > \lambda_c$: The dynamical spin polarization (a)$\langle\langle \sigma_x \rangle\rangle_r$, (c) $\langle\langle \sigma_y \rangle\rangle_r$, (e) $\langle\langle \sigma_z \rangle\rangle_r$ on dBIS at $\boldsymbol{k}_c$. $\boldsymbol{k}_c$ (red stars) denotes the singularity momentum (defined in the main text or the right plane of this figure), the spin polarization dynamics first lose oscillation on the dBIS at $\boldsymbol{k}_c$ with the increase of noise leading to ill-defined dBIS and topological trivial phases. The time averaged spin polarization (b) $\overline{\langle\langle \sigma_x \rangle\rangle}_r$, (d) $\overline{\langle\langle \sigma_y \rangle\rangle_r}$, (f) $\overline{\langle\langle \sigma_z\rangle\rangle_r}$ at the momentum $k_y=0$.  The purple lines (Ideal) show the results of a clean system. The grey lines (M) show the results under Markovian noise. The blue lines from lightest to heaviest (NM1,NM2,NM3,NM4) show the results of non-Markovian noise with non-Markovianity parameter $b = 2, 1, 0.6, 0.2$. The system parameters are chosen as $m_z=1.2,t_{so}=0.2,t_0=1$.
} \label{fig:2Dstrong}   % \label{}
\end{figure}

\begin{figure}[!htb]
\includegraphics[width=0.95\textwidth]{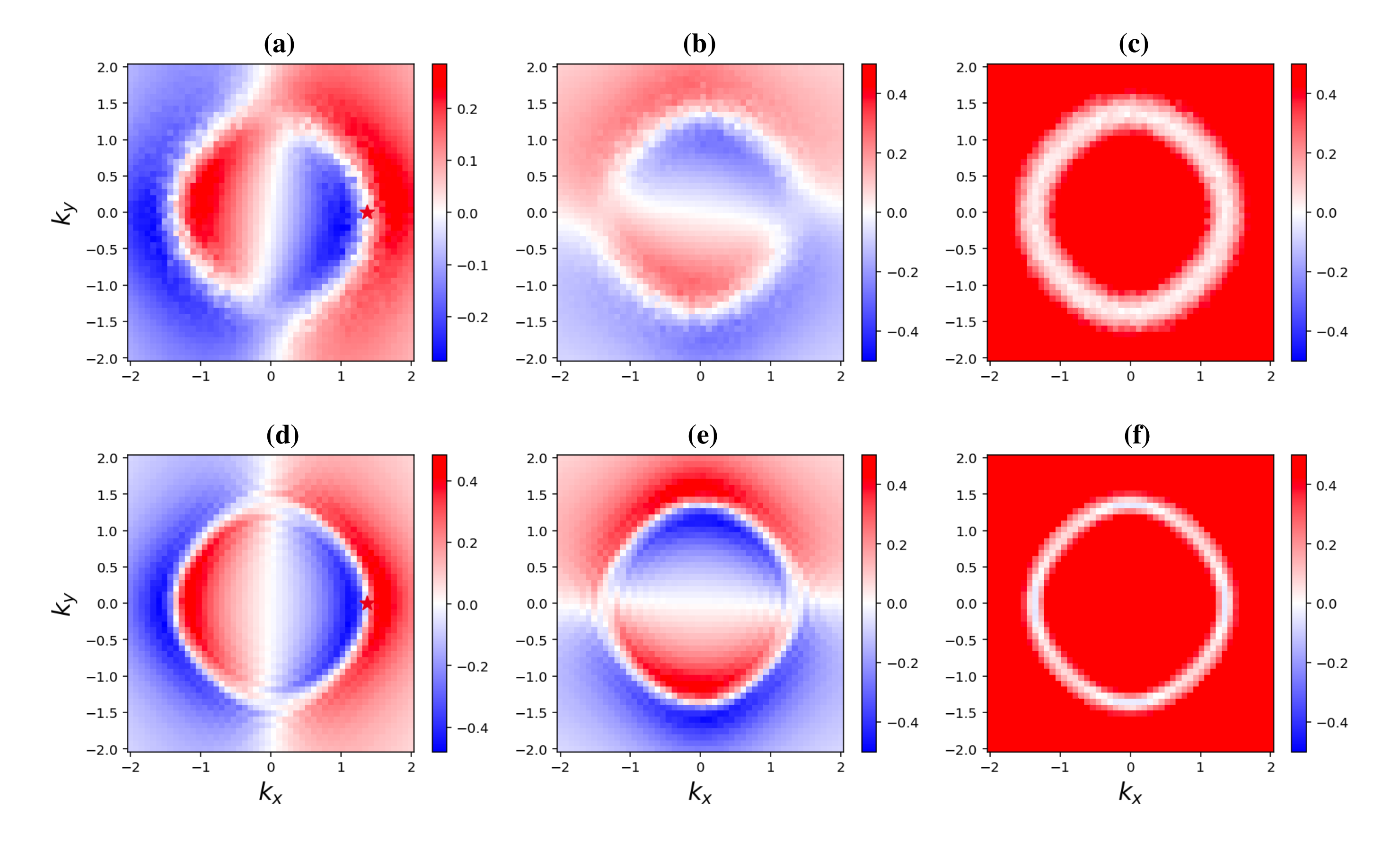}
\caption{Time-averaged spin polarization under weak noise $\lambda=0.2$ in the Markovian limit:  (a) $\overline{\langle\langle \sigma_x \rangle\rangle_r}$, (b) $\overline{\langle\langle \sigma_y \rangle\rangle_r}$, (c) $\overline{\langle\langle \sigma_z \rangle\rangle_r}$. Time-averaged spin polarization under weak noise $\lambda=0.2$ with non-Markovianity parameter $b=0.2$:  (d) $\overline{\langle\langle \sigma_x \rangle\rangle_r}$, (e) $\overline{\langle\langle \sigma_y \rangle\rangle_r}$, (f) $\overline{\langle\langle \sigma_z \rangle\rangle_r}$.
} \label{fig:3Dweak}   % \label{}
\end{figure}

\begin{figure}[!htb]
\includegraphics[width=0.95\textwidth]{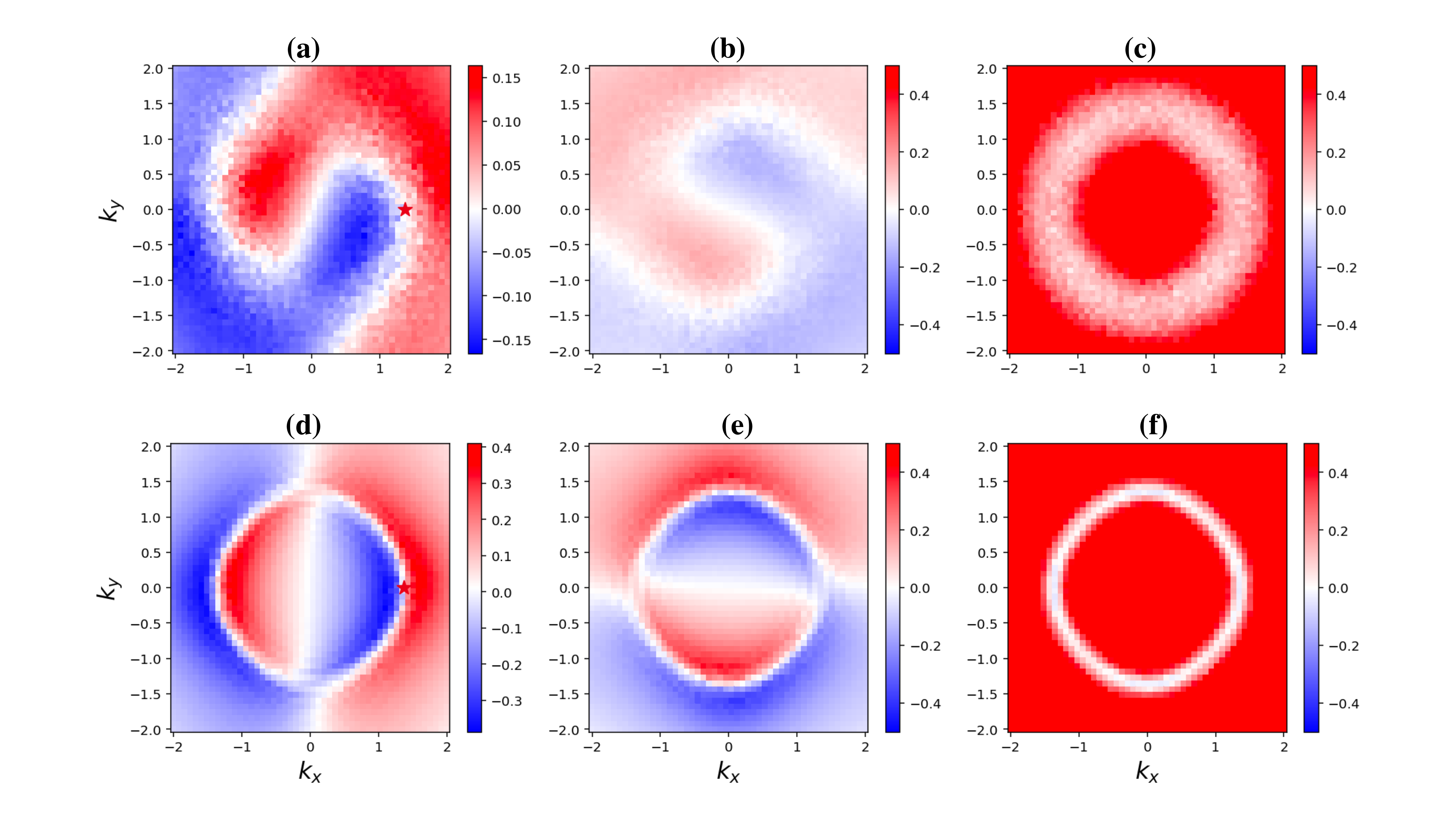}
\caption{Time-averaged spin polarization under strong noise $\lambda=0.8$ in the Markovian limit:  (a) $\overline{\langle\langle \sigma_x \rangle\rangle}_r$, (b) $\overline{\langle\langle \sigma_y \rangle\rangle}_r$, (c) $\overline{\langle\langle \sigma_z \rangle\rangle}_r$. Time-averaged spin polarization under strong noise $\lambda=0.8$ with non-Markovianity parameter $b=0.2$:  (d) $\overline{\langle\langle \sigma_x \rangle\rangle}_r$, (e) $\overline{\langle\langle \sigma_y \rangle\rangle}_r$, (f) $\overline{\langle\langle \sigma_z \rangle\rangle}_r$. 
} \label{fig:3Dstrong}   % \label{}
\end{figure}

In this section, we summarize the theory of quench-induced dynamical topology of the 2D quantum anomalous Hall (QAH) mode with Markovian noise and present the simulation details and results with the influence from non-Markovianity.

Extensive studies have been carried out in recent years on topological quantum matter \cite{hasan2010colloquium,qi2011topological,chiu2016classification, sato2017topological}, surpassing the established Landau-Ginzburg-Wilson framework \cite{landau1999statistical}. Notably, researchers have discovered that topological phases, originally defined in the ground state under equilibrium conditions, can also be explored using quantum quenches, which provide a nonequilibrium approach to studying topological physics \cite{vajna2015topological,caio2015quantum,budich2016dynamical,wilson2016remnant,flaschner2018observation,unal2020topological,hu2020topological}. In this context, the concept of the dynamical bulk-surface correspondence has been introduced as a momentum-space counterpart to the bulk-boundary correspondence. This correspondence establishes a connection between the bulk topology of an equilibrium phase and the emergence of a nontrivial dynamical topological phase on specific momentum subspaces known as band-inversion surfaces (BISs) when the system undergoes a quench across topological transitions \cite{zhang2018dynamical,zhang2019dynamical,zhang2020unified,zhang2022universal}. This dynamical topology offers a versatile method for characterizing and detecting topological phases through quantum dynamics, leading to numerous experimental studies in various quantum simulation platforms, including ultracold atoms \cite{sun2018uncover,yi2019observing}, nuclear magnetic resonance (NMR), nitrogen-vacancy defects in diamond \cite{ji2020quantum, xin2020quantum}, and superconducting circuits \cite{niu2021simulation}.
 Once it comes to simulation experiments on quantum platforms, environmental noise becomes an inevitable problem. Recently, there has been a surge of interest in exploring the interplay between system topology and noise environment, where the influence has led to the emergence of diverse dynamical topological phenomena~\cite{heiss2012physics,lee2016anomalous,leykam2017edge,xu2017weyl,kunst2018biorthogonal,yao2018non,zhang2021quench,lin2022experimental}.

According to ref.~\cite{zhang2021quench}, the stochastic Markovian noise with $\langle \langle B^\alpha(t) \rangle\rangle_{r}=0$ and $\langle\langle B^\alpha(t)B^\beta(t') \rangle\rangle_{r}=\lambda_\alpha \delta(t-t')\delta_{\alpha,\beta}$ make the quantum dynamics of the QAH system follow the master equation
\begin{equation}
\frac{d\rho(t)}{dt}=-i[H_{QAH},\rho(t)]+\sum_{\alpha=x,y,z}B_\alpha[\sigma_\alpha\rho(t)\sigma_\alpha-\rho(t)].
\end{equation}
The dynamical spin polarization is defined by $\mathbf{s}(\boldsymbol{k},t)\equiv \langle\langle \boldsymbol{\sigma}(\mathbf{k},t)\rangle\rangle_r=\text{Tr}[\rho(\mathbf{k},t)\boldsymbol{\sigma}]$, $\rho(\mathbf{k},t)$ is the density matrix which can be expressed as $\rho(\mathbf{k},t)=1/2[1+\mathbf{s}(\mathbf{k},t)\cdot\boldsymbol{\sigma}]$ for the two band model. The dynamical evolution of spin polarization is governed by Liouvillian superoperator as $\partial_t \mathbf{s}(\mathbf{k},t)=\mathcal{L}(\mathbf{k})\mathbf{s}(\mathbf{k},t)$.
 For the two-band model, the Liouvillian superoperator can be recast into the following matrix $\mathcal{L}(\mathbf{k})$ 
\begin{equation}
\mathcal{L}(\mathbf{k})=2
\begin{bmatrix}
-B^y-B^z & -h_z(\mathbf{k}) & h_y(\mathbf{k}) \\
h_z(\mathbf{k}) & -B^x-B^z & -h_x(\mathbf{k})  \\
-h_y(\mathbf{k}) & h_x(\mathbf{k}) & -B^x-B^y  \\
\end{bmatrix}.
\label{eq:liouv}
\end{equation}
By diagonalizing the Liouvillian superoperator, the dynamical spin polarization can be explicitly acquired
\begin{equation}\label{eq:st}
\mathbf{s}(\mathbf{k},t)=\mathbf{s_0}(\mathbf{k})e^{-\gamma_0(\mathbf{k})t}+\mathbf{s_+}(\mathbf{k})e^{-[\gamma_1(\mathbf{k})+i\omega(\mathbf{k})]t}+\mathbf{s_-}(\mathbf{k})e^{-[\gamma_1(\mathbf{k})-i\omega(\mathbf{k})]t},
\end{equation}
where the coefficients $\mathbf{s}_i(\mathbf{k})=(\mathbf{s}_i^L(\mathbf{k})\cdot\mathbf{s}(\mathbf{k},0))\mathbf{s}_i^R(\mathbf{k})$ for $i=0,\pm$. Here $\mathbf{s}_i^{L(R)}$ are the left(right) eigenvectors of the Liouvillian superoperator 
\begin{equation}
\mathcal{L}^T(\mathbf{k})\mathbf{s}_i^L=-\gamma_i\mathbf{s}_i^L,\ \mathcal{L}(\mathbf{k})\mathbf{s}_i^R=-\gamma_i\mathbf{s}_i^R,
\end{equation}
with eigenvalues $\gamma_0$ and $\gamma_\pm=\gamma_1\pm i\omega$ respectively. $\omega$ denotes the oscillation frequency of the dynamical spin polarization. $\gamma_{0,1}$ explicitly reflects the spin dynamics dissipation due to the noise.  

The original noiseless dynamical topology is defined on BIS, which denotes a momentum subspace $\{\mathbf{k}|h_z(\mathbf{k})=0\}$ with $\mathbf{h}(\mathbf{k})\cdot \mathbf{s}(\mathbf{k},0)=0$, where the spin-flip resonant oscillations occur and thus the time-averaged spin-polarization vanishes: $\lim_{T\rightarrow{\infty}}\int_0^T dt \  \mathbf{s}(\mathbf{k},t)|_{\mathbf{k}\in BIS}=0$. The noise deformed BIS(dBIS) under noise can also be well defined 
\begin{equation}
\text{dBIS}\equiv\{\mathbf{k}|\mathbf{s}_0^L\cdot\mathbf{s}(\mathbf{k},0)=0\},
\end{equation}
if only the noise strength is restricted in the so-called sweet spot region
\begin{equation} \label{eq:fullregion}
\max_{\mathbf{k}\in \text{dBISs}}[(\lambda_y-\lambda_x)h_x^2/\mathbf{h}_{so}^2-2|\mathbf{h}_{so}|]<\lambda_z-\lambda_x<\min_{\mathbf{k}\in \text{dBISs}}[(\lambda_y-\lambda_x)h_x^2/\mathbf{h}_{so}^2+2|\mathbf{h}_{so}|],
\end{equation}
where $\mathbf{h}_{so}(\mathbf{k})=(h_x,h_y)$. For the initial state $\vert\uparrow\rangle$, the dBIS can be analytically given by the momenta satisfying $h_z=(\lambda_y-\lambda_x)h_xh_y/(h_x^2+h_y^2)$ with $\mathbf{s}_0^L\sim(h_x,h_y,0)$. For weak noise in the sweet spot region, the noise fails to close the bulk gap of the
emergent dynamical phase, the dynamical spin polarization retains finite oscillation behavior in the entire Brillouin zone. The corresponding noisy dynamical spin polarization in Eq.~\eqref{eq:st} can be rescaled to the one in the clean system by neglecting the amplitude decay while keeping the oscillation
\begin{equation}
\mathbf{s}'(\mathbf{k},t)\equiv\mathbf{s_0}(\mathbf{k})+\mathbf{s}_+(\mathbf{k})e^{-i\omega(\mathbf{k})t}+\mathbf{s}_-(\mathbf{k})e^{i\omega(\mathbf{k})t}.
\end{equation}
Thus, there is an emergent topology of quench dynamics characterized by the topological winding number,
\begin{equation}
\mathcal{W}=\frac{1}{2\pi}\oint_{\text{dBIS}}\mathbf{g}(\mathbf{k})d\mathbf{g}(\mathbf{k}),
\end{equation}
where $\mathbf{g}(\mathbf{k})$ denotes the normalized gradient of averaged dynamics spin polarization on dBIS.

For sufficient strong noise outside the sweet spot region indicated by Eq.~\eqref{eq:fullregion}, the noisy spin polarization only exhibits decay behavior without oscillation in time. Thus it cannot be rescaled to the one as in the clean system by neglecting the amplitude decay $\mathbf{s}'(\mathbf{k},t)=\sum_{i=0,\pm}\mathbf{s}_i(\mathbf{k})=\mathbf{s}(\mathbf{k},0)$. Under this situation, noise induces singularity on the dBIS, rendering ill defined dBIS. The quench dynamics thus belongs to topological trivial phases. 

We next focus on the pure dephasing bath defined in Sec.~\ref{sec:appnmpure} to study the non-Markovian effect on the quench-induced dynamical topology. 
Based on the definition of dBIS, dBIS is equal to BIS in the noiseless limit under pure dephasing case ($\lambda_x=\lambda_y=0$, $\lambda_z=\lambda$) with initial state $|\uparrow\rangle$. In the Markovian limit, the sweet point region defined in Eq.~\eqref{eq:fullregion} is reduced as 
\begin{equation} \label{eq:region1}
\lambda \leq \lambda_c= \min_{\mathbf{k}\in \text{dBISs}} 2\|(h_x(\mathbf{k}),h_y(\mathbf{k}))\|.
\end{equation}
$\lambda_c$ here is given by $\mathbf{k}_c$, which denotes the singularity momentum.  The spin polarization dynamics first lose oscillation at $\mathbf{k}_c$  on the dBIS with the increasing noise strength, leading to ill-defined dBIS and the topological trivial phase.

What will happen when non-Markovianity is introduced?  We present the dynamical simulation results from Fig.~\ref{fig:2Dweak} to Fig.~\ref{fig:3Dstrong}. 

For weak noise in the sweet spot region (Fig.~\ref{fig:2Dweak}(a),(c),(e)), the dynamical spin polarization at $\boldsymbol{k}_c$ exhibits decay companing with oscillation behavior in the Markovian limit. After rescaling the dynamical process by neglecting the amplitude decay while keeping the oscillation, dBIS can be well defined ($\lim_{T\rightarrow{\infty}}\int_0^T dt \  
\mathbf{s}(\mathbf{k},t)|_{\mathbf{k}\in dBIS}=0$) and the dynamical topology is preserved. Once we turn on non-Markovianity gradually, the dynamical spin polarization automatically approaches clean cases with weaker amplitude decay compared to the original Markovian case. The same conclusion can be drawn from the time-averaged spin polarization for $k_y=0$ (Fig.~\ref{fig:2Dweak}(b),(d),(f)) and in the whole Brillouin zone (Fig.~\ref{fig:3Dweak}). 

For strong noise strength outside the sweet spot region (Fig.~\ref{fig:2Dstrong}(a),(c),(e)), the dynamical spin polarization at $\boldsymbol{k}_c$ exhibits serious decay without well-defined oscillation behavior (oscillation frequency peak is $\omega=0$ given by Fourier transformation) in the Markovian limit. Therefore, dBIS is ill defined and it is topological trivial. Once the non-Markovianity is turned on gradually, the dynamical spin polarization exhibits similar behavior as the weak noise case with finite oscillation frequency approaching the noiseless case. Thus dBIS is well defined and the topology is successfully recovered from original harmful noise thanks to non-Markovianity.  The same conclusion can be drawn from the time-averaged spin polarization for $k_y=0$ (Fig.~\ref{fig:2Dstrong}(b),(d),(f)) and in the whole Brillouin zone (Fig.~\ref{fig:3Dstrong}). We note that, to give an intuitive perspective of the impact caused by noise strength and non-Markovianity,  all the results presented are directly from numerical simulation without any error mitigation rescaling operations, i.e., we keep the amplitude decay induced by dephasing.

All the findings support that the presence of non-Markovianity in noise enhances the dynamics simulation process and facilitates the transition of a trivial phase into a topological phase in the presence of strong noise.

\section{Non-Markovianity in Many-body Localization Dynamics} \label{sec:appmbl}

\begin{figure}[!htb]
\includegraphics[width=0.5\textwidth]{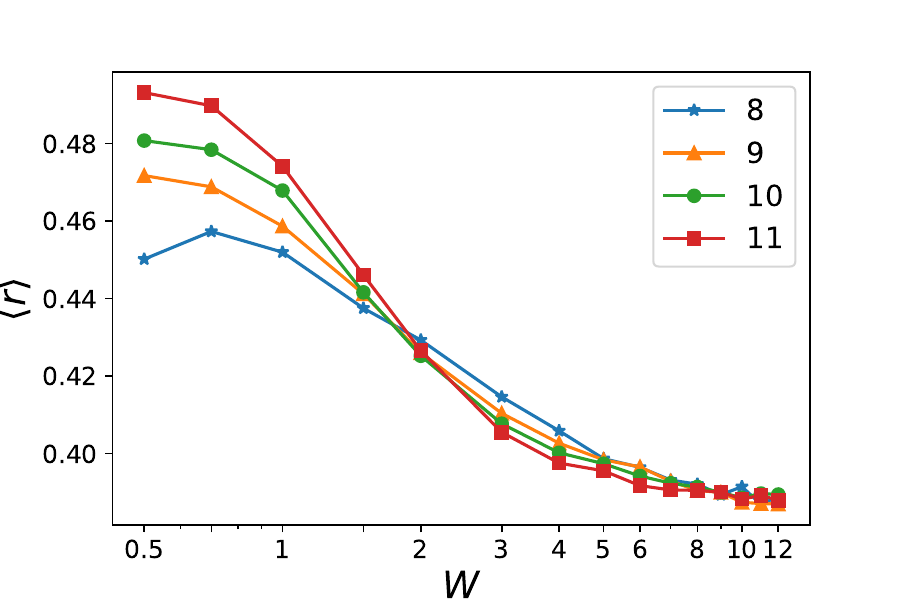}
\caption{The average ratio of adjacent energy gaps. The system size $N$ is given in the legend. In the thermal phase, the system has Gaussian-orthogonal ensemble (GOE) level statistics while in the many-body localized phase the level statistics follow Poisson distribution. In the main text, we set $W=10$ to make the system deep in the localized phase.} \label{fig:mbl_phase}   % \label{}
\end{figure}

\begin{figure}[!htb]
\includegraphics[width=0.95\textwidth]{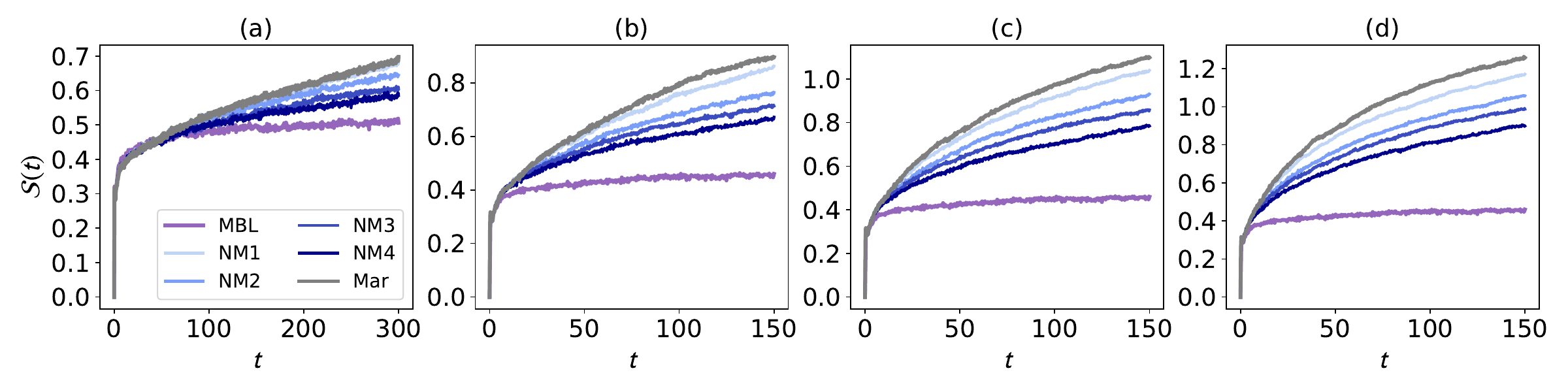}
\caption{Entanglement entropy of many-body localization dynamics with pure dephasing noise of strength $\lambda= 0.3$(a), $\lambda = 0.5$(b), $\lambda = 0.6$(c), $\lambda = 0.8$(d).
The purple line shows the results of the clean MBL system. The grey line shows the results under Markovian noise. Blue lines labeled from NM1 to NM4 show the results under non-Markovian noise with non-Markovianity parameter
$b = 4, 1, 0.5, 0.2$.} \label{fig:sm_mbl_ee}   % \label{}
\end{figure}

\begin{figure}[!htb]
\includegraphics[width=0.95\textwidth]{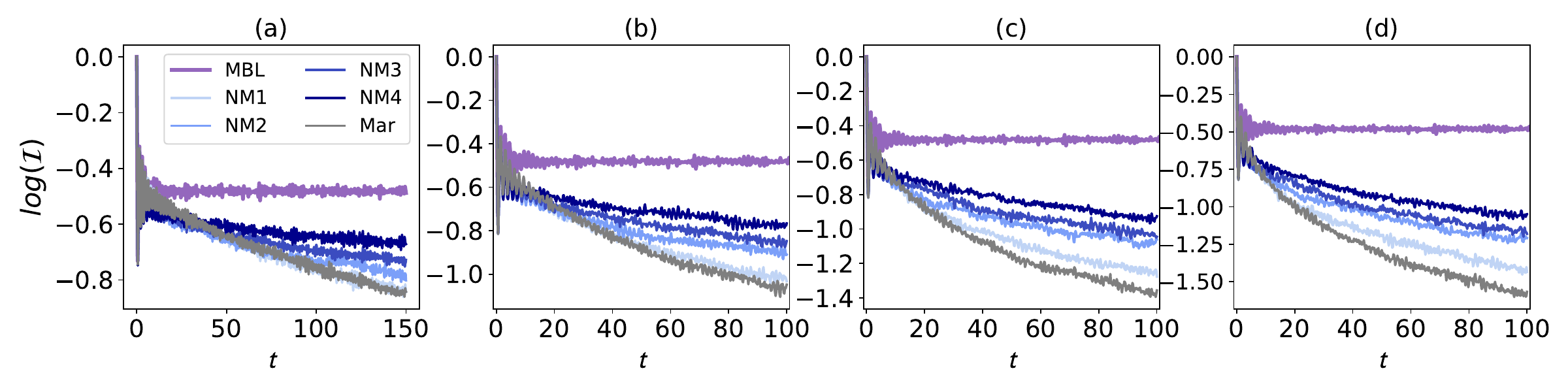}
\caption{Charge imbalance of many-body localization dynamics with pure dephasing noise of strength $\lambda= 0.3$(a), $\lambda = 0.5$(b), $\lambda = 0.8$(c), $\lambda = 1.0$(d).
The purple line shows the results of the clean MBL system. The grey line shows the results under Markovian noise. Blue lines labeled from NM1 to NM4 show the results under non-Markovian noise with non-Markovianity parameter $b =
4, 1, 0.5, 0.2$. 
} \label{fig:sm_mbl_i}   % \label{}
\end{figure}

As explained in the main text, we use the ``standard model” for MBL -- a one dimensional $S=1/2$ spin chain with nearest neighbor interaction and strong random disorder for on-site potentials - to investigate the dynamics of MBL with and without quantum noise. Note that to be compatible with dephasing noise distribution, the on-site random potential of MBL is chosen to follow the Gaussian distribution $N(0,W/\sqrt{3})$. This distribution gives the same mean and variance values for random variables $W_i$ compared to the uniform distribution $U(-W, W)$.

We first study the static property of this model to determine the phase boundary between MBL and thermal phases in terms of the disorder strength $W$. As shown in Fig.~\ref{fig:mbl_phase}, we compute the averaged level spacing ratio of the eigenspectrum of the system. The quantity is defined as $\langle r\rangle = \langle \frac{\min\{\delta_n, \delta_{n+1}\}}{\max\{\delta_n, \delta_{n+1}\}}\rangle$, where the average $\langle \cdot\rangle$ is defined over different energy level and different disorder configurations, $\delta_n=e_{n+1}-e_{n}$ is the adjacent level spacing for eigenstates $e_n$. In the MBL phase, the level spacing of the spectrum is expected to follow Possion distribution with no level repulsion $\langle r\rangle \approx 0.386$, while in the thermal phase, the level spacing follows Wigner-Dyson distribution $\langle r\rangle\approx 0.53$. The crossing point of $\langle r\rangle$ of different system sizes corresponds to the delocalization transition point, which is around $W_c\approx 2$ in our case. In the dynamical simulation, we set $W=10$ to make the system deeply in the MBL phase.

As explained in the main text, we use entanglement entropy $S$ and the charge imbalance $\mathcal{I}$ to characterize the dynamics. The initial state for the quench dynamics is the $Z_2$ product state $\vert 0101....\rangle$. Suppose the stochastically evolved state for a given noise trajectory at the time $t$ is $\vert \psi(t)\rangle$, charge imbalance is defined as
\begin{equation}
\mathcal{I}(t) = \sum_i^N (-1)^i \langle \psi(t)\vert  S^z_i \vert \psi(t)\rangle,
\end{equation}
And the quantity can be further averaged over different disorder realizations.

The half-chain reduced matrix is defined as the partial trace over half of the spin freedom on the evolved pure state under given disorder configurations as $\rho_A(t) = \text{Tr}_B (\vert \psi(t)\rangle \langle \psi(t)\vert)$, where A and B stand for each half chain of the system. The entanglement entropy is thus defined as:
\begin{equation}
S(t) = - \text{Tr}(\rho_A(t)\ln \rho_A(t)).
\end{equation}
The quantity can be further averaged over different disorder realizations. Note the order here cannot be exchanged due to the non-linear nature of the entropy definition. %The entropy of disorder averaged evolved density matrix will give trivial thermal value.

Under quantum dephasing noise of different strengths with and without non-Markovianity, we numerically simulate the dynamics using stochastic processes with each time slice $\delta=0.1$ and $500$ disorder configurations. We extract the asymptotic behavior of the two average quantities as summarized in Fig.~\ref{fig:sm_mbl_ee} and Fig.~\ref{fig:sm_mbl_i}.

Note that the charge imbalance is a linear quantity with respect to the state $\rho$, so the computation order (average over different MBL disorder configurations, average over different noise space-time profiles, compute the charge imbalance) doesn't matter. However, the entanglement entropy depends on the state $\rho$ in a non-linear fashion, so that the computation order in the numerical simulation matters. If we first average over different quantum noise profiles, the resulting state is a mixed state and the entanglement entropy will contain a large part from the thermal fluctuation, failing to capture the intrinsic entanglement. Though in this case, we can utilize other probes such as mutual information or entanglement negativity to determine the intrinsic entanglement components from the averaged mixed state. Instead, we compute the entanglement entropy for each trajectory (each disorder configuration and quantum noise profile), so that the entanglement is all from pure state contribution, reflecting the intrinsic property of the eigenstates of MBL system. This subtle difference in computation order also exists in measurement induced entanglement phase transition studies.

\end{document}

\title{The quantum imaginary-time control for accelerating the ground-state preparation}% Force line breaks with \\

\author{Yu-Cheng Chen}
\affiliation{Tencent Quantum Laboratory, Tencent, Shenzhen, Guangdong, China, 518057}
\affiliation{Department of Mechanical Engineering, City University of Hong Kong, Hong Kong SAR, 999077, China}
\author{Yu-Qin Chen}
\email{yuqinchen@tencent.com}
\affiliation{Tencent Quantum Laboratory, Tencent, Shenzhen, Guangdong, China, 518057}
\author{Alice Hu}
\email{alicehu@cityu.edu.hk}
\affiliation{Department of Mechanical Engineering, City University of Hong Kong, Hong Kong SAR, 999077, China}
\affiliation{Department of Materials Science and Engineering, City University of Hong Kong, Kowloon, Hong Kong SAR, 999077, China}
\author{Chang-Yu Hsieh}
\email{kimhsieh2@gmail.com}
\affiliation{Tencent Quantum Laboratory, Tencent, Shenzhen, Guangdong, China, 518057}
\affiliation{{Innovation Institute for Artificial Intelligence in Medicine of Zhejiang University, College of Pharmaceutical Sciences, Zhejiang University, Hangzhou, 310058, China}}
\author{Shengyu Zhang}
\affiliation{Tencent Quantum Laboratory, Tencent, Shenzhen, Guangdong, China, 518057}
\maketitle
\onecolumngrid
\appendix

\tableofcontents

\section{Acceleration Induced by Lyapunov Control} \label{apptheory}
In this section, we explain the simulation acceleration driven by Lyapunov control in multiple stages. First, we discuss a practical strategy to construct a control Hamiltonian (without explicitly diagonalizing the system Hamiltonian) that ensures substantial speedup in section \ref{appgeneralcontrol}. Second, we attempt to provide an intuitive understanding of the accelerated dynamical process by inspecting time-dependent changes in the distribution of eigenstate populations and the time-evolved energy spectrum of $H(\tau)$ in section \ref{appdynamicpross}. In section \ref{appcontrol}, we look at a broad spectrum of realistic models (of interest in condensed matter physics, quantum chemistry, and combinatorial optimization) to demonstrate the broad applicability of the proposed method for simulating the ground state of a Hamiltonian.

\subsection{General Strategy For Control Hamiltonian Construction} \label{appgeneralcontrol}

As stated in the previous section, one scenario to drive the desired exponential speedup is to construct an $H_d$ that (1) commutes with $H_p$, (2) maintains the eigenstate ordering of $H_p$ and (3) holds the matrix norm for $H_p + H_d$ roughly the same as that of $H_p$. Among the three criteria, it is reasonable to partially relax the condition on the eigenstate ordering and make $H_d(t)$ a time-dependent control Hamiltonian instead.

One possibility is to choose $H_d(t)=\sum_{i=1}^n \beta_i(t) H_p^i$, which is an $n$-th order matrix polynomial made up of power of $H_p^i$ and $\beta_i \in \mathbf{R}$. Intuitively, the time-dependent spectrum of $H_d(t)$ may temporarily alter eigenstate ordering and enlarge or shrink energy gaps between states. Under this temporal modulation of the Hamiltonian spectrum, there would be accelerated and de-accelerated population transfer among eigenstates of $H_p$ during the imaginary time evolution. To ensure that we obtain as large an energy filtering towards the ground state of $H_p$ as possible is to simply minimize the expected energy value $\bra{\psi(t)}H_p\ket{\psi(t)}$ during the imaginary time evolution under $H_p+H_d(t)$. The gradient descent will give us the rule to update $\beta(t)$. At this point, this strategy of optimizing $H_d(t)$ essentially corresponds to the closed-loop control theory discussed thoroughly in the main text.

We now have a simple and clear picture for the selection of control Hamiltonian. However, it might not be experimentally feasible to implement such a complex form of control Hamiltonian, $H_d(t)=\sum_{i=1}^n \beta_i(t) H_p^i$. An approximate option is to retain only the major components of $H_p$ in this polynomial expansion of $H_d(t)$. Effectively, instead of having an $H_d(t)$ that always commutes with $H_p$, we end up considering an $\tilde H_d(t)$ with non-zero off-diagonal matrix elements in the eigenbasis of $H_p$. While this compromise (due to realistic experimental constraints) may offset the exponential speedup, we find that it is still possible to achieve an appreciable speedup within the closed-loop control setup. 
To make the discussion in this section self-contained, we present a numerical illustration.

In Figure \ref{Dtest} we show the simulation result of how the imperfectly controlled dynamics, driven by $H_p + \tilde H_d(t) =  U \tilde{D}(t) U^\dag$, affect the convergence of imaginary-time evolution for a  9 qubits 2D XXX model with non-periodic boundary condition. The problem Hamiltonian can be written as $H_p=0.2\sum_i Z^i+0.1 \sum_{edge}X^iX^j+Y^iY^j+Z^iZ^j$, where edge implies the nearest-neighbor couplings and the initial state is $\ket{++\dots+}$. Assume that the originally intended control Hamiltonian $H_d$ is intended to enlarge the energy gaps between the ground state and excited states, but, due to the experimental constraints, we choose to adopt a more easily implementable $\tilde H_d$. In this case, we no longer have a fully diagonal matrix $D$ in the eigenbasis of $H_p$ rather we consider 
$$\tilde{D}=\begin{pmatrix}
-5& 0 & 0 & ...\\
0 & 5 & 0 & ...\\
... & ... & ... & ...\\
0 & 0 & 0 & 0
\end{pmatrix}+\mathcal{R}(p),$$
where $\mathcal{R}(p)$ is a random sparse matrix with $p$ percentage of non-zero off-diagonal matrix elements. The result shows that the $H_d$ with non-trivial off-diagonal disorder $\tilde{D}(p=10\%)$ can still maintain the exponential speed-up for this problem, although the further increase of the number of off-diagonal terms will eventually suppress the speedup. Interestingly, when the number of non-zero off-diagonal disorders rises up to $p=60\%$ the inhibition effects on the speedup of this 9-qubit system seem to plateau. 

In a practical scenario, the control Hamiltonian for the models in Appendix \ref{appcontrol} follows the rules we mentioned above. For molecular systems studied, Pauli Z terms are the dominant terms in $H_p$ with large coefficients. The selection of  those dominant Pauli $Z$ terms as control Hamiltonian follows the rule (1) since it commutes with the highest contribute terms in $H_p$ therefore approximately commutes with $H_p$. The details of $H_d$ selection are listed in Appendix \ref{appmoleculecontrol}. For complex systems like the XXX model and SK model, Pauli $Z$ terms from $H_p$ may not be dominant parts. We select the $k$-local Pauli $Z$ as $H_d$, which not only commutes with all Pauli $Z$ terms but also commutes with many Pauli XX(YY) terms making it approximately follow the rule (1). Finally, proper design of the control strategy and maximum admissible control impulses will allow rules (2) and (3) to be successfully followed, see details in \ref{appstrategy}.

\begin{figure*}
\begin{center}
\hspace*{-0cm}\includegraphics[width=0.8\textwidth]{FigureS1.pdf}
\caption{The convergence results of $H_d$. The ITC matrix density is the percentage of the matrix $\tilde{D}$ filled with the nonzero number range from 0 to 1. The x-axis is the log of the total convergence time step. The y-axis is the fidelity between the convergence state and the ground state.
}\label{Dtest}
\end{center}
\end{figure*}

%=========================================
\subsection{Dynamic process visualization} \label{appdynamicpross}

We find the ITC temporally modulates the entire spectrum of $H(\tau)$ during the evolution and return back to the original spectrum of $H_p$ when time $t>>1$ as $\beta(t>>1)\rightarrow 0$. As can see in the following figures, for the case of the imaginary time evolution (ITE), since the $H_p$ is a time-independent Hamiltonian, the energy gap between the ground state and the excited states remains fixed in time. For the imaginary time control (ITC), on the other hand, the energy gap between the ground state and the excited states of $H_p$ will change with time. However, unlike the polynomial basis that can guarantee to order of the whole eigenspectrum as in Figure \ref{FULLH2}, the small set of limited Pauli basis might only modify some of the eigenstates as in Figure \ref{PARTH2}. We comment on how the energy gap changes with time under control using different control Hamiltonians from the point of view of controllability, which provides another realization of how the control Hamiltonian orders the energy spectrum during the convergence. The idea is borrowed from the idea of controllability discussed in real-time control \cite{Zhang2005}, in our definition, given a set of reachable states denoted by $\mathcal{R}(\ket{\psi_0})$ start from initial state $\ket{\psi_0}$, the control system $H_p+\beta(\tau) H_d$ is said to be completely (eigenstates) controllable if $\mathcal{R}(\ket{\psi_0})=\mathcal{E}$, where $\mathcal{E}$ is the collection of eigenstates, and said to be non-completely (eigenstates) controllable if $\mathcal{R}(\ket{\psi_0})\neq\mathcal{E}$. From the point of view of the energy spectrum, since the final convergence state of imaginary time evolution should be the ground state of the new system, the control system should reorder the whole energy spectrum during the convergence when the control system is completely controllable. For the control system that is not completely controllable, it can still accelerate the ground state convergence if the ground state in the reachable state $\mathcal{R}(\ket{\psi_0})$ which should always be true since ITC can switch back to the original ITE by the turn of the $\beta$. By definition, it is clear that polynomial basis is in the class of completely controllable, and limited Pauli basis might be in one of the classes according to the elements in the basis. In the following numerical experiment, we will visualize the concept of controllability using the energy spectrum.
The energy gap between the ground state and the excited states of $H_p$ will change with time as,
$$
\Delta E(\tau) = \expval{H(\tau)}{\psi_0}-\expval{H(\tau)}{\psi_i}.
$$
where $H(\tau)=H_p+\beta(\tau)H_d$, $\ket{\psi_0}$ and $\ket{\psi_i}$ correspond to the ground state and i-th excited state of drift Hamiltonian $H_p$, respectively. In Fig.~\ref{FULLH2} we show the simulated result of using an ITC to prepare the ground state for a 2-qubit hydrogen model. In this case, we adopt a set of control Hamiltonians that yield complete controllability (polynomial basis), on the 2-qubit dynamics. As clearly illustrated in the figure, the ITC temporally modulates the entire spectrum of $H(\tau)$ during the evolution and returns back to the original spectrum of $H_p$ when time $t>>1$ as $\beta(t>>1)\rightarrow 0$.  Thus, we conclude that the controlled system can evolve towards the ground state of $H_p$ in fewer time steps with enlarged energy gaps.

Next, we attempt a different experiment. If the control Hamiltonian does not guarantee complete controllability(limited Pauli basis), the imaginary-time control will not be able to enlarge all energy gaps but just a few of them. In this case, it can still speed up the convergence of the ITC but may not be as fast as the previous case with complete controllability In Fig.~\ref{PARTH2}, the 2-qubit Hydrogen is driven by a non-complete controllable Hamiltonian $H_d$, the instantaneous ground state energy drop increases the energy gap between the ground state and the excited states, therefore, provide the speed-up for ITC, and this control will benefit more when the initial state has an appreciable overlap with the highest excited state since the highest excited state is also controllable. 
\par
\begin{figure}[ht]
\begin{center}
\includegraphics[width=0.7\textwidth,height=0.4\textwidth]{FigureS2.pdf}
\caption{The result of 2-qubit hydrogen with complete control. The left figure shows how the energy levels change with time, as the solid line is the original energy level and the dashed line is the controlled energy level. The right figure is the convergence of imaginary time models. The left y-axis corresponds to the energy for the red solid line and blue solid line. The right y-axis corresponds to the energy gap between the ground state and the first excited state, and the $\beta(\tau)=\frac{1}{n}\sum_{i=1}^n\beta_i(\tau)$. The result shows that all energy levels will change with time.}\label{FULLH2}
\end{center}
\end{figure}

\par
\begin{figure}[ht]
\begin{center}
\includegraphics[width=0.7\textwidth,height=0.4\textwidth]{FigureS3.pdf}
\caption{The result of 2-qubit hydrogen with non-complete control shows that the highest excited state and the ground state will both change with time.}\label{PARTH2}
\end{center}
\end{figure}
\par
However, when the control Hamiltonian is not completely controllable and the control strategy $\beta(\tau)$ is too aggressive, then the driven ITE may actually compromise the rate of convergence. This is because the ground state has been inverted to the highest excited state in some time steps, and the system will be temporarily driven away from the target state (i.e. the ground state of $H_p$) when its energy has been shifted upwards. In Fig.~\ref{badH2}, during the
period $\tau \in (0.5,1.0)$, the order of the ground state and the highest excited state are swapped. Clearly, as shown in the right panel of Fig.~\ref{badH2}, the ITE converges to the desired ground state faster in this case.
\begin{figure}[ht]
\begin{center}
\includegraphics[width=0.7\textwidth,height=0.4\textwidth]{FigureS4.pdf}
\caption{The result of 2-qubit hydrogen with aggressive control strategy shows that the control pulses are able to rearrange energy level.}\label{badH2}
\end{center}
\end{figure}

\begin{figure}[ht]
\begin{center}
\includegraphics[width=0.7\textwidth,height=0.4\textwidth]{FigureS5.pdf}
\caption{The result of 2-qubit hydrogen with different admissible maximum strength S. It shows that the large admissible maximum strength might decrease the energy gap.}\label{S1VS10}
\end{center}
\end{figure}

To avoid the unintended eigenstate re-ordering by the control field, it is crucial to confine the magnitude of $\beta$ to some proper range. In our test, it is better to set $\vert \beta \vert$ to be equal to or less than the magnitude of the energy $\bra{\psi(\tau)}H_p \ket{\psi(\tau)}$ which can be obtained from measuring the norm of the matrix problem Hamiltonian. In Fig.~\ref{S1VS10} we illustrate the effects of choosing different $\beta$ on the time-evolved energy levels of $H_p$, which is the Hamiltonian for a 2-qubit Hydrogen. In this test, we adopt two different approximate bang-bang control strategies $\beta(\tau)=\frac{2S}{1+e^{-\gamma T_k}}-S$, introduced in appendix \ref{appstrategy}. These two control strategies share the same $\gamma$ but $S=1$ and $S=10$, respectively. From Fig.~\ref{S1VS10}, it is clear that bigger $\vert \beta\vert$ may enlarge energy gaps between eigenstates of $H_P$ but also increases the likelihood of state re-ordering.

\section{Empirical Control Hamiltonian Selection} \label{appcontrol}
As mentioned in Appendix \ref{apptheory}, the success of the imaginary time control (ITC) is to modify this typical evolution path of ITE. The control may drive the time-evolved wavefunction to temporarily enhance contributions from high-energy eigenstates before converging to the ground state. In other words, the advantage of having controlled dynamics in the imaginary-time domain is to avoid densely spaced energy regions in the Hamiltonian spectrum, which cannot be escaped in the standard ITE. If one can find a $H_d$ that could drive a system towards high-energy states during the ITE, it has the potential to improve the convergence efficiency as prescribed by our proposed method to get ITC. 

In this section,  we present the strategies of control Hamiltonian selection and discuss how different control Hamiltonian impact the 
the execution of quantum computing experiments, including empirically control Hamiltonian and simulation results for molecule in Appendix \ref{appmoleculecontrol}, control Hamiltonian, and control Hamiltonian and simulation results for Sherrington–Kirkpatrick in Appendix \ref{appskcontrol}, control Hamiltonian and simulation results for a spin model constructed from 3-SAT in Appendix \ref{appsatcontrol}.
In summary, with a proper design of control Hamiltonian, we will obtain obvious acceleration for  ground state simulation. Thus our proposed imaginary-time control can potentially make a digital quantum simulation for the ground-state preparation much more accessible in the NISQ era.

\subsection{Control Hamiltonian and simulation results for molecule system} \label{appmoleculecontrol}
Based on the above general strategy for control Hamiltonian selection, here we provide a further empirical method for control Hamiltonian selection, especially for molecule systems. We find that the empirical $H_d$ and its slightly modified versions can exponentially speed up the ITC convergence for studied systems with both large and small energy gaps.

For the molecule system, we try the candidate control Hamiltonian $H_i$ containing only Pauli $Z$ (single $Z$ and double $Z$, and the total number of choices is $C_1^n+C_2^n$ where $n$ is a number of qubits. In studied molecular cases, we find that limiting the Pauli $Z$ to quadratic is enough to obtain exponential speed-up. 

Intra-molecular bond distance for the typical dissociation curves is considered as an adjustable parameter to generate variant energy gaps for the molecular systems. We choose a problem Hamiltonian $H_p$ with a large energy gap as a fast test to help with control Hamiltonian selection. We first run standard ITE on this $H_p$ (which should converge to the ground state easily owing to the large energy gap) and keep a record of the $\beta_i(\tau)$ from the ITE calculation with $\ket{+++ \dots ++}$ as the initial state. We sum up $\beta_i$ over time $B_i=\sum_{\tau} \beta_i(\tau)$ for further analysis. From our numerical investigations on molecular systems: H chain, LiH, and HF, we find that those $H_i$ having negative $B_i$ all share the same structure. The structures of Hamiltonian means that, the Pauli strings are divided into two groups that occupy $[0:(N-1)]$ and $[N:(M-1)]$ orbitals respectively, where $N$ is the number of electrons and $M$ is the number of total orbitals. We use these $H_i$ as the control Hamiltonian, which we called \emph{Empirical $H_d$}. Empirical $H_d$ and its slightly modified versions can exponentially speed up the ITC convergence for studied systems with both large and small energy gaps.

 \begin{table}[!htb]
\begin{tabular}{|l|c|c|c|c|c|c|}
\hline
 &
  \multicolumn{1}{l|}{$H_2$} &
  \multicolumn{1}{l|}{$H_4$} &
  \multicolumn{1}{l|}{$H_6$} &
  \multicolumn{1}{l|}{$H_8$} &
  \multicolumn{1}{l|}{LiH} &
  \multicolumn{1}{l|}{HF} \\ \hline
\multirow{4}{*}{Full $H_d$} &
  II+P(IZ) &
  IIII+P(IIIZ) &
  IIIIII+P(IIIIIZ) &
  IIIIIIII+P(IIIIIIIZ) &
  IIII+P(IIIIIIIZ) &
  IIIIIIIIII+P(IZ) \\ \cline{2-7} 
 & P(IZ)+II & P(IIIZ)+IIII & P(IIIIIZ)+IIIIII & P(IIIIIIIZ)+IIIIIIII & P(IIIZ)+IIIIIIII & P(IIIIIIIIIZ)+II \\ \cline{2-7} 
 & II+P(ZZ) & IIII+P(IIZZ) & IIIIII+P(IIIIZZ) & IIIIIIII+P(IIIIIIZZ) & IIII+P(IIIIIIZZ) & IIIIIIIIII+P(ZZ) \\ \cline{2-7} 
 & P(ZZ)+II & P(IIZZ)+IIII & P(IIIIZZ)+IIIIII & P(IIIIIIZZ)+IIIIIIII & P(IIZZ)+IIIIIIII & P(IIIIIIIIZZ)+II \\ \hline
\multirow{2}{*}{Half $H_d$} &
  II+P(IZ) &
  IIII+P(IIIZ) &
  IIIIII+P(IIIIIZ) &
  IIIIIIII+P(IIIIIIIZ) &
  IIII+P(IIIIIIIZ) &
  IIIIIIIIII+P(IZ) \\ \cline{2-7} 
 & II+P(ZZ) & IIII+P(IIZZ) & IIIIII+P(IIIIZZ) & IIIIIIII+P(IIIIIIZZ) & IIII+P(IIIIIIZZ) & IIIIIIIIII+P(ZZ) \\ \hline
 \multirow{2}{*}{All $H_d$} &
  P(IIIZ) &
  P(IIIIIIIZ) &
  P(IIIIIIIIIIIZ) &
  P(IIIIIIIIIIIIIIIZ) &
  P(IIIIIIIIIIIZ) &
  P(IIIIIIIIIIIZ) \\ \cline{2-7} 
 & P(IIZZ) & P(IIIIIIZZ) & P(IIIIIIIIIIZZ) & P(IIIIIIIIIIIIIIZZ) & P(IIIIIIIIIIZZ) & P(IIIIIIIIIIZZ) \\ \hline
\multirow{3}{*}{Empirical $H_d$} &
  II+P(IZ) &
  IIII+P(IIIZ) &
  IIIIII+P(IIIIIZ) &
  IIIIIIII+P(IIIIIIIZ) &
  IIII+P(IIIIIIIZ) &
  IIIIIIIIII+P(IZ) \\ \cline{2-7} 
 & II+P(ZZ) & IIII+P(IIZZ) & IIIIII+P(IIIIZZ) & IIIIIIII+P(IIIIIIZZ) & IIII+P(IIIIIIZZ) & IIIIIIIIII+P(ZZ) \\ \cline{2-7} 
 & P(ZZ)+II & P(IIZZ)+IIII & P(IIIIZZ)+IIIIII & P(IIIIIIZZ)+IIIIIIII & P(IIZZ)+IIIIIIII & P(IIIIIIIIZZ)+II \\ \hline
\end{tabular}
\caption{The $H_d$ selection of figures. The function P(S) is the collection of all permutations without repetition of Pauli string S. The I+P(S) means add I before all the Pauli string P(S), for example, II+P(IZ) is equal to \{IIIZ, IIZI\}. The Full $H_d$ is slightly different from empirical $H_d$ since it has the P(IZ)+I term and the Half $H_d$ is slightly different from the empirical $H_d$  since it lacks the P(ZZ)+I term.
}\label{T1}
\end{table}
Take it for an example, there are 4 electrons ($N=4$) and 12 total orbitals ($M=12$), the Pauli strings $P_0P_1P_2P_3P_4P_5P_6P_7P_8P_9P_{10}P_{11}$ will be divided into two groups: $P_0 P_1 P_2 P_3 IIIIIIII$ and $IIIIP_4 P_5 P_6 P_7 P_8 P_9 P_{10} P_{11}$ (for Openfermion notation, the energy of the orbitals from low to high are ordered from left to right). We find that the single $Z$ Pauli strings that $Z$ only exist in $[N:(M-1)]$ orbitals and the double $Z$ Pauli strings that $Z$s exist only in $[0:(N-1)]$ orbitals or $[N:(M-1)]$ orbitals have negative $B_i$ (as Table \ref{T1} ``Empirical $H_d$"), and all other Pauli strings have positive $B_i$. 

Finally, we use the empirical $H_d$ described above for various molecules (H chain, LiH, and HF, see Appendix \ref{appmolecularham}.) with different energy gaps. In Figure \ref{HD1} and Figure \ref{HD2}, we show the ITC results with the control Hamiltonian that has all single $Z$ and double $Z$s (``control all" in Figure \ref{HD1} and Figure \ref{HD2}, ``All Hd" in Table \ref{T1}), the empirical control Hamiltonian mentioned above (``control negative" in Figure \ref{HD1} and Figure \ref{HD2}, ``Empirical Hd" in Table \ref{T1}) and the result of the standard ITE ("no control" in Figure \ref{HD1} and Figure \ref{HD2}). The results show that the empirical $H_d$ is far better than using all $H_i$ as control Hamiltonian. And empirical $H_d$ can exponentially speed up the ITC convergence for studied systems with both large and small energy gaps.

\begin{figure*}
\begin{center}
\hspace*{-0cm}\includegraphics[width=1\textwidth]{FigureS6.pdf}
\caption{The result of the control Hamiltonian test of the H chain system, the x-axis is a log of 1/$\Delta E$, and the y-axis is the total time steps for convergence. The result shows that the empirical control Hamiltonian can provide an ITC that gives exponential speed-up with respect to the standard ITE for all cases while $H_d$ that contain all $H_i$ require extra time or no speed-up.
}\label{HD1}
\end{center}
\end{figure*}
\begin{figure*}
\begin{center}
\hspace*{-0cm}\includegraphics[width=0.95\textwidth]{FigureS7.pdf}
\caption{The result of the control Hamiltonian test of LiH and HF, the x-axis is a log of 1/$\Delta E$, and the y-axis is the total time steps for convergence. The result shows that the empirical control Hamiltonian can provide good control results that have exponentially speed-up from ITE for all cases while $H_d$ that contain all $H_i$ require extra time or no speed-up.
}\label{HD2}
\end{center}
\end{figure*}
\begin{figure*}
\begin{center}
\hspace*{-0cm}\includegraphics[width=0.9\textwidth]{FigureS8.pdf}
\caption{The result of the H chain size scaling test with different types of $H_d$. The different dots represent different types of $H_d$ and different colors of lines represent the total time steps.  
}\label{Hlog}
\end{center}
\end{figure*}

We also test two types of $H_d$ that are slightly different from the empirical $H_d$ mentioned above. The details of those $H_d$ are also listed in Table \ref{T1}. Figure \ref{Hlog} shows the convergence of different types of $H_d$ selections and the speed-up for different system sizes, where the ``Full $H_d$" have extra single $Z$ terms compared to ``Empirical $H_d$" and the ``Half $H_d$" have less double Z terms compared to ``Empirical $H_d$". We can see that for ``Full $H_d$" and ``Half $H_d$" the acceleration scale exponentially as a function of $1/\Delta E$ for different choices of $H_p$. 

\subsection{Control Hamiltonian and simulation results for Sherrington–Kirkpatrick model} \label{appskcontrol}
Besides the commute basis $\{XX\dots XX, YY\dots YY, ZZ\dots ZZ\}$ we use in the 4 qubit cases in the main text. We also construct and compare two types of $H_d$ according to Appendix \ref{appgeneralcontrol}. One is constructed polynomial $H_d$ from the approximate polynomial basis $p(\tilde{H}_p)=\{\tilde{H}_p^2,\tilde{H}_p^3,\tilde{H}_p^4,\tilde{H}_p^5\})$, where approximate polynomial matrix $\tilde{H}_p^k$ is constructed from polynomial matrix $H_p^k$ by removing matrix elements that have values less than a fixed constant number. The other Pauli $H_d$ is selected from the 1-local Pauli Z and 4-local Pauli Z $cyclic(ZIIIIII)$ and $cyclic(ZZZZIIII)$, where cyclic means the set of circular shifts of Pauli strings, for example, $cyclic(XYZ)=\{XYZ, YZX, ZXY\}$.
The control strategy we use here is approximate bang-bang control with the maximum strength of the control field S slowly turning to zero within dozens of time steps to have lower resource requirement, see Appendix \ref{appstrategy}. 

\begin{figure}[ht]
\begin{center}
\includegraphics[width=0.7\textwidth]{FigureS9.pdf}
\caption{The x-axis is the case indices sorted by the ratio of total QITE time steps divided by the total QITE time steps, note that sorting does not depend on energy differences. The y-axis is the ratio of the total time step. The mean of $p(H)$ and $\sigma(H)$ is the mean of the approximate polynomial $H_d$ and Pauli $H_d$ of ten randomly selected initial states. The best of $p(H)$ and $\sigma(H)$ is the largest speed-up among ten randomly selected initial states using approximate polynomial $H_d$ and Pauli $H_d$. The single time step $\Delta\tau=0.03$.}\label{SK}
\end{center}
\end{figure}
In Figure \ref{SK}, we compared the convergence of ITE and ITC of 100 randomly sampling cases and each case has 10 randomly selected initial states. The result shows that the control can provide a general speed-up and can achieve a hundred times speed-up in some cases. The approximate polynomial $H_d$ can provide a higher average speed-up but weaker corner case speed-up, while Pauli $H_d$ has a lower average speed-up but has a better corner case speed-up.

\FloatBarrier
\subsection{Control Hamiltonian and simulation results for a spin model constructed from 3-SAT} \label{appsatcontrol}
In the main text, we show that ITC can provide obvious speed-up over the ITE for molecular systems. It is also desirable to verify whether such a superior advantage can hold for other scenarios. In this subsection, we compare ITC and ITE for solving a spin model that is closely related to the 3-SAT problems. This is  another ground-state preparation task that is sufficiently distinct from the molecular systems considered in the main text. 

3-SAT problem is defined by a logical statement involving $n$ boolean variables $b_i$. The logical statement consists of m clauses $C_i$ in conjunction:
$C_1 \wedge C_2 \wedge ... \wedge C_m$. Each clause is a disjunction of 3 literals, where a literal is a boolean variable $b_i$ or its negation $\lnot b_i$. For instance, a clause may read $(b_j \lor b_k \lor b_l).$ The task is to first decide whether a given 3-SAT problem is satisfiable; if so, then assign appropriate binary values to satisfy the logical statement. We can map a 3-SAT problem to a Hamiltonian for a set of qubits. Under this mapping, each binary variable $b_i$ is represented as a qubit state. Thus, an n-variable 3-SAT problem is mapped into a Hilbert space of dimension $N = 2^n$. Furthermore, each clause of the logical statement is translated to a projector acting on this n-qubit system. Hence, a logical statement with m clauses may be translated to the following
Hamiltonian, 

$$
H_{final}=\sum_{\alpha = 1}^m\ket{b_j^\alpha b_k^\alpha blj^\alpha}\bra{b_j^\alpha b_k^\alpha blj^\alpha}.
$$
A common approach to solve this type of constraint satisfaction problem is to use adiabatic quantum computations(AQC). One first prepares the ground states of an easy-to-solve Hamiltonian $H_{init}$. Next, one slowly evolves the Hamiltonian such that it adiabatically connects $H_{init}$ and $H_{final}$. In other words, the adiabatically evolved Hamiltonian reads
$$
H(s)=(1-s)H_{init}+sH_{final}, s \in [0,1],
$$
where $H_{init}$ is typically chosen to be a sum of one-qubit Hamiltonians $H_i$ acting on
the i-th qubit,
$$
H_{init}=\frac{1}{2}\sum_{i=1}^{n}h_i,\quad h_i=\begin{pmatrix}
    1 & -1 \\
   -1 & 1
\end{pmatrix}.
$$
The energy gap $\Delta E(s)$ between the instantaneous ground state and first-excited state of $H(s)$ will vary with time $s$
as shown in Fig~.\ref{dEdS}. Based on the well-established theoretical studies, it is also clear that the instantaneous energy gap of interest can get very small along the adiabatic path when the system size is large.

\begin{figure}[ht]
\begin{center}
\includegraphics[width=0.6\textwidth]{FigureS10.pdf}
\caption{The energy gap between the ground state and first excited state of the time-dependent Hamiltonian $H(s).$}\label{dEdS}
\end{center}
\end{figure}

\FloatBarrier
\section{The imaginary time Control Strategy} \label{appstrategy}
For the imaginary-time control, in this letter, we mainly used two types of control strategy. The first one is the approximate bang-bang control introduced above that target the larger energy gap problem (type I). The second one is the modified version of the bang-bang strategy that targets the small energy gap problem (type II). We also provide the source codes of all simulations at \url{https://github.com/Kesson-C/QITC}, where you can find more details on the control strategy and control parameter settings.
\subsection{The imaginary time Control Strategy- Type I} \label{appsatistrategyI}
For the type I control strategy, let us generalize the standard Lyapunov control such that it could work with the imaginary-time evolution. We redefine $T_k\equiv 2\expval{H_p}{\psi}\expval{H_d}{\psi}-\expval{\{H_p,H_d\}}{\psi}$ in this case. The $H_d$ related terms in $T_k(\tau)$ may entail lots of extra measurements if they cannot be obtained by measuring the Pauli terms appearing in $\expval{H_p}{\psi}$. To reduce the measurement cost and still maintain a powerful $H_d$ to provide an enhanced convergence, we propose the following strategy. We first decide if the state in the quantum circuit has high overlap with any eigenstate of $H_p$ or $H_d$ by checking the value of $T_k$, if $T_k < L$ then we do not apply any control pulses, otherwise we use a similar control strategy for the real-time case introduced above.
$$\beta_k(\tau)=\left\{
\begin{array}{rcl}
&\frac{2S}{1+e^{-\gamma T_k}}-S,&  (T_k\geq L)\\
&0,&  else
\end{array} \right.
$$
where L is some pre-defined threshold value. If the state is close to an eigenstate (i.e. $T_k(\tau)<L$), we should turn off the control field and let the system evolves under $H_p$ in the imaginary-time domain. This truncation can greatly reduce the measurement costs (for the implementation of the corresponding variational algorithm) in the region where the state will linearly converge to the eigenstate. Finally, we test how the truncation (i.e. setting $\beta_k(\tau)=0$ when $T_k(\tau)<L$) will affect the precision of the converged results given by truncating the control pulse in the middle of the convergence (Figure \ref{50step}). For a physical system with a large energy gap between the first-excited state and the ground state, it only requires a few control steps at the beginning then the result of convergence will be very close to the result without truncation. For systems with a small energy gap, the truncation will significantly affect the precision of the converged results.
\begin{figure}[h]
\begin{center}
\includegraphics[width=0.9\textwidth]{FigureS11.pdf}
\caption{The numerical control term truncation test result of 8 qubits HF molecule with different energy gap}\label{50step}
\end{center}
\end{figure}
\subsection{The imaginary time Control Strategy- Type II} \label{appsatistrategyII}
For the type II control strategy, we use a modified bang-bang control strategy that has two phases. The first phase of control will bypass the slow convergence region and the second phase will speed up the convergence, the demonstration of the phases is demonstrated in Figure \ref{2phase}. For the small energy gap problem, the type I control might drift the system to the low excited state and weaken the speed-up. To avoid this, we control the state to the transition state that contains high energy excited state and then to the ground state in two phases. Phase I of the control will lower the contribution of the low-energy states, the energy will go up and reach the equilibrium state consisting of high-energy excited states. Phase two is the traditional bang-bang control which is designed to lower the energy. The switch from phase I to phase II is decided by the energy changes (reach the equilibrium state), and the time-energy plot of the type two control is shown in Figure \ref{2phase}. In summary, we propose the following control strategy,
$$\beta_k(\tau)=\left\{
\begin{array}{rcl}
&K_1sgn(T_k(\tau)),&  (phase 1)\\
&-K_2sgn(T_k(\tau)),&  (phase 2)
\end{array} \right.
$$
where $K_1$ and $K_2$ are the strength of the control field and $sgn(T_k(\tau))$ is the sign of the $T_k(\tau)$, $\beta_k(\tau)$ will be set to zero if the sign alternating between positive and negative(reach the bound of the $H_d$), in this letter we set $K_1>>K_2$ to avoid convergence to the excited state.
\begin{figure}[h]
\begin{center}
\includegraphics[width=0.9\textwidth]{FigureS12.pdf}
\caption{Two-phase control example, the convergence will pass through three different phases. At phase I, the control will drift the state to a higher energy transition state. In phase II, the control will speed up the convergence to the ground state. In the final phase, the control will be closed to save computing resources.}\label{2phase}
\end{center}
\end{figure}
In conclusion, the advantage of the type I control strategy has weak requirements for the selection of control Hamiltonian, but it may not provide exponential speed-up of the 1/$\Delta E$. The type II control strategy on the other hand can provide exponentially speed-up of the 1/$\Delta E$ but require stronger control Hamiltonian selection.

\FloatBarrier
\section{Molecular Hamiltonian} \label{appmolecularham}
\subsection{Hydrogen}
In our simulations, we consider the hydrogen molecule in the minimal STO-3G basis. Each
The hydrogen atom contributes a single 1S orbital. As a result of the spin, there are four spin-orbitals in total.
By using the function of the Qiskit, the qubit Hamiltonian for JW representation can be obtained. This 4-qubit Hamiltonian is given by:
\begin{align*}
H&=h_0I+h_1Z_0+h_2Z_1+h_3Z_2+h_4Z_3+h_5Z_0Z_1\\
&+h_6Z_0Z_2+h_7Z_0Z_3+h_8Z_1Z_2+h_9Z_1Z_3+h_{10}Z_2Z_3\\
&+h_{11}Y_0Y_1X_2X_3+h_{12}Y_0Y_1Y_2Y_3+h_{13}X_0X_1X_2X_3\\
&+h_{14}X_0X_1Y_2Y_3+h_{15}X_0X_1Y_2Y_3
\end{align*}
By using the function of the Qiskit, the 2-qubit Hamiltonian of the hydrogen molecule can be obtained from parity representation with Z2 symmetry reduction. This 2-qubit Hamiltonian is given by:
\begin{align*}
H&=h_0I+h_1Z_0+h_2Z_1+h_3Z_1Z_0+h_4X_1X_0
\end{align*}
And the circuit (Figure \ref{cvqe}) for Variational-based ansatz\cite{Sim2019} simulation used in the main text.
\begin{figure}[h]
\begin{center}

\includegraphics[width=0.8\textwidth]{FigureS13.pdf}
\caption{The variational circuit of 4 qubit $H_2$ system}\label{cvqe}
\end{center}
\end{figure}
\par
To update the parameters of Variational-based ansatz, we calculate the gradient of the circuit by  using parameter shift rule\cite{Schuld2019} to obtain the numerical differential result of the circuit.

\subsection{Hydrogen Chain}
In our simulations, we consider the Hydrogen chain molecule in the minimal STO-3G basis. The Hydrogen Chain $H_n$ has n electrons and contributes n 1S orbital to the basis.

\subsection{Hydrogen Fluoride}
In our simulations, we consider the Hydrogen fluoride molecule in the minimal STO-3G basis. The Fluoride atom has 9 electrons, and so contributes a 1S, 2S, $2P_x$, $2P_y$, and $2P_z$ orbital to the basis, while the Hydrogen atom contributes a single 1S orbital. For the appendix B truncation test, by freezing the core two 1S orbitals, we can reduce the system from 12 orbitals with 10 electrons to 8 orbitals with 6 electrons on 2S and 2P orbitals. By using the Qiskit, the qubit Hamiltonian for Jordan-Wigner representation can be obtained. This 8-qubit Hamiltonian is given by 145 different Pauli terms and coefficients.

\subsection{Lithium Hydride}
In our simulations, we consider the Lithium hydride molecule in the minimal STO-3G basis. The Lithium atom has 3 electrons, and so contributes a 1S, 2S, $2P_x$, $2P_y$, and $2P_z$ orbital to the basis, while the Hydrogen atom contributes a single 1S orbital.